\documentclass[a4paper]{jpconf}

\usepackage{graphicx}

\graphicspath{{./figures/}}

\begin{document}
\title{Higgs physics: Review of recent results and prospects from ATLAS and CMS}

\author{Martin Flechl on behalf of the ATLAS and CMS collaborations}

\address{Institute of High Energy Physics, Austrian Academy of Sciences, Wohllebengasse 12-14, 1040 Vienna, Austria}

\ead{martin.flechl@cern.ch}


\setlength{\textfloatsep}{17pt}

\begin{abstract}
An overview of recent results in Higgs boson physics obtained with the ATLAS and CMS experiments at the Large Hadron Collider located at CERN, Geneva, 
is presented. The focus is on measurements of the properties of the recently discovered Higgs boson with a mass of about 125 GeV. A brief selection of results 
in searches for Higgs bosons beyond the Standard Model is given, and prospects of future Higgs boson measurements and searches at the LHC are discussed.
\end{abstract}

\section{Introduction}
In July 2012, the ATLAS~\cite{atlas} and CMS~\cite{cms} collaborations announced the discovery of a new particle compatible with being the Higgs boson of the Standard 
Model (SM)~\cite{disc_atlas,disc_cms}. Subsequent measurements of the properties of this particle are all consistent with the SM Higgs boson interpretation. 
In the following sections, the most recent results from studies of this Higgs boson using the data collected with the ATLAS and CMS experiments are presented. 
First, the results in various Higgs boson production and decay channels are given, followed by the measurement of Higgs boson properties using a combination of 
these results. 
Searches for Higgs bosons predicted by theories beyond the Standard Model (BSM) are briefly summarized and prospects for future Higgs boson property 
measurements are discussed. Most of the results are based on the full data set recorded in 2011 and 2012 at center-of-mass energies of 7 TeV and 8 TeV, respectively, 
corresponding to about 25 fb$^{-1}$ per experiment.

\section{Higgs boson search results}\label{sec:chan}
The dominant Higgs boson production modes are gluon-gluon fusion (ggF), 
vector boson fusion (VBF), production in association with a vector boson $V$ ($VH$, $V=W$ or $V=Z$), 
and in association with top quarks ($ttH$). 
The five most sensitive Higgs boson decay channels at the LHC are 
the modes $\gamma\gamma$, 4-leptons, $WW$, $\tau\tau$ 
and $b\bar{b}$~\cite{yr1,yr2,yr3}. Results for these channels are 
summarized in the following.

\subsection{$H \to \gamma\gamma$}
\begin{figure}[htb]
\begin{center}
\begin{minipage}{0.40\textwidth}
\includegraphics[height=6.3cm]{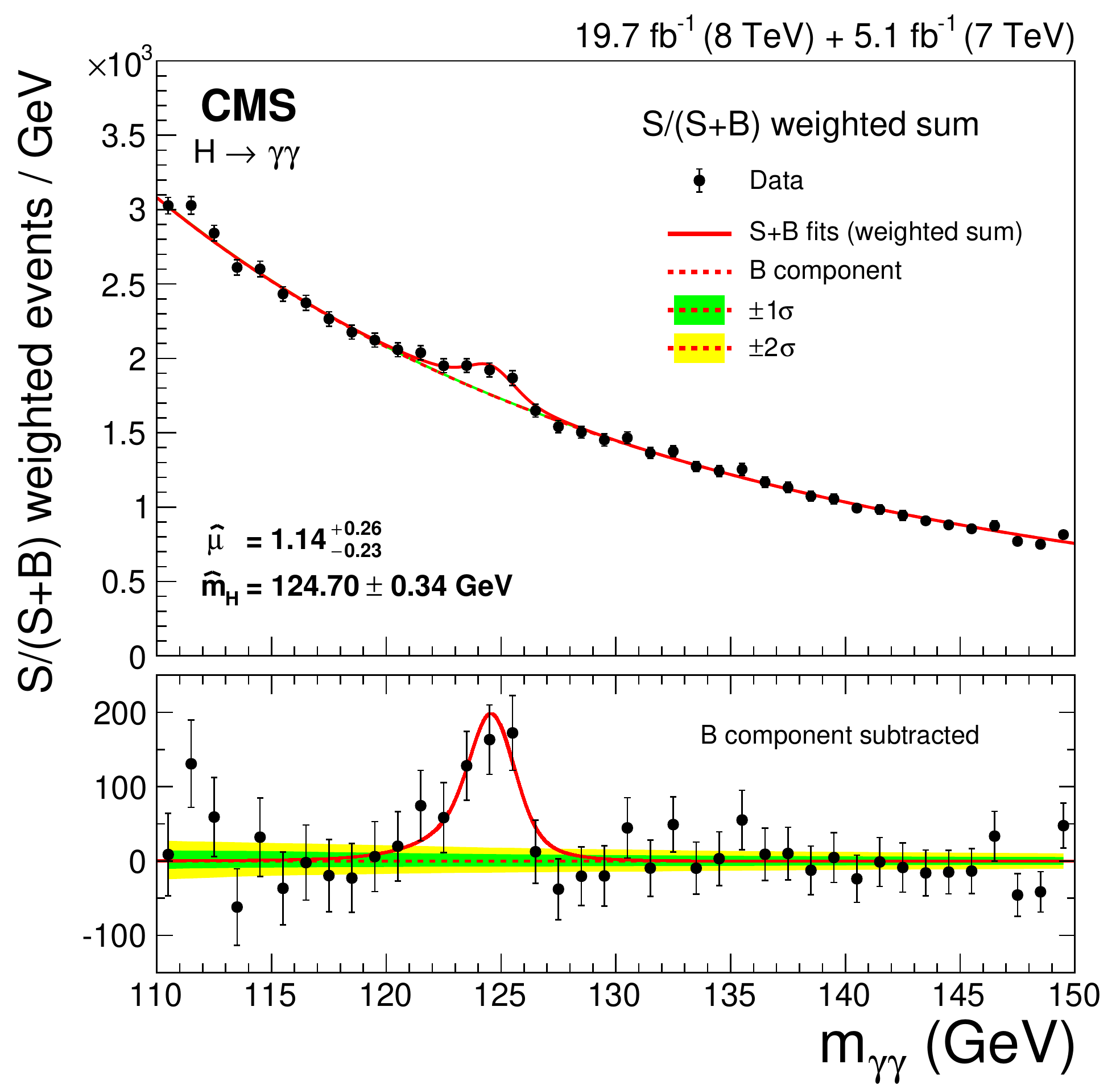}
\caption{\label{hgg_c1}The CMS $m_{\gamma\gamma}$ distribution as weighted sum of all categories~\cite{hgg_cms}. $S$ and 
$B$ are the number of signal and background events in a small mass window around $m_H$ for each event category, respectively.} 
\end{minipage}\hspace{0.05\textwidth}%
\begin{minipage}{0.53\textwidth}
\includegraphics[height=6.3cm]{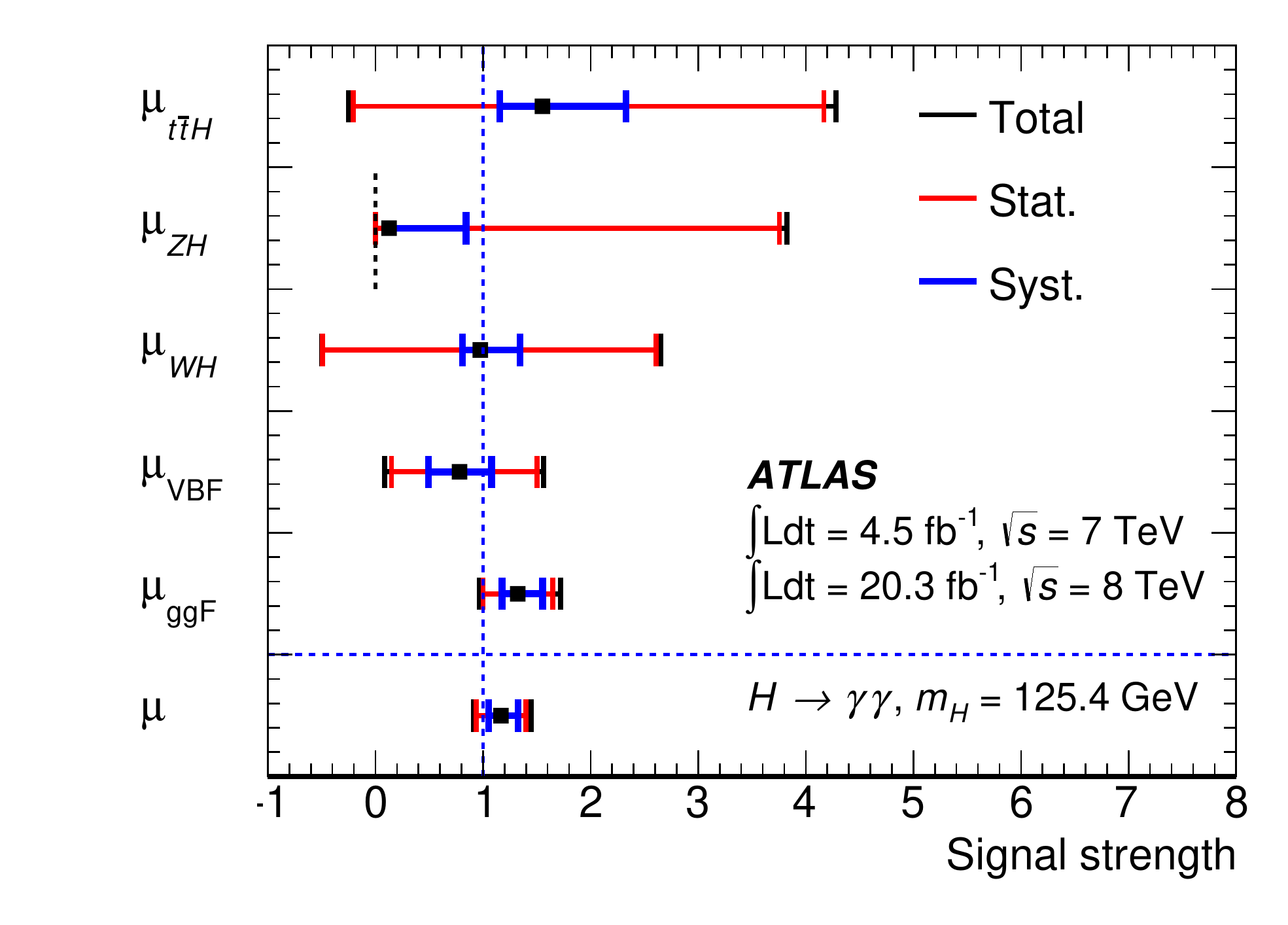}
\caption{\label{hgg_a1}Signal strength in the $H \to \gamma\gamma$ channel. Shown is the measurement for the individual Higgs production modes 
as well as the combined ATLAS value~\cite{hgg_atlas}. }
\end{minipage}
\end{center}
\end{figure}
The $H \to \gamma\gamma$ channel is characterized by relatively high total event counts but a low signal-over-background ratio. 
Furthermore, it offers a high mass resolution ($m_{\gamma\gamma}$) and thus contributes strongly to the Higgs boson mass 
measurement. The analyses~\cite{hgg_cms,hgg_atlas} proceed by selecting events with two photons in various categories, aimed at different production 
modes. This is done by additionally requiring two jets with a high rapidity gap (VBF), additional leptons and in some cases 
missing energy ($VH$) or an event topology consistent with an additional top quark pair event ($ttH$). The final discriminating variable 
is the $m_{\gamma\gamma}$ estimator. The background, dominantly continuum $\gamma\gamma$, $\gamma$+jet and di-jet events, is estimated by 
fitting the sidebands of the $m_{\gamma\gamma}$ distribution, as illustrated in Fig.~\ref{hgg_c1}. The signal is visible on top of the estimated 
background at $m_{\gamma\gamma} \approx 125$~GeV.

Both ATLAS and CMS observe slightly more $H \to \gamma\gamma$ candidate events than expected; however, the measurement is within 
one Gaussian standard deviation ($\sigma$) of the SM expectation. Both experiments measure the signal strength $\mu$ (observed cross section times 
branching ratio divided by the SM expectation) also individually for the various Higgs boson production modes, see Fig.~\ref{hgg_a1}. 
All values are in agreement with the SM expectation of $\mu=1$. The observed signal significance is $5.7\sigma$ (expected: $5.2\sigma$) for 
CMS and $5.2\sigma$ (expected: $4.6\sigma$) for ATLAS.

\subsection{$H \to 4l$}
\begin{figure}[htb]
\begin{center}
\begin{minipage}{0.47\textwidth}
\includegraphics[height=7cm]{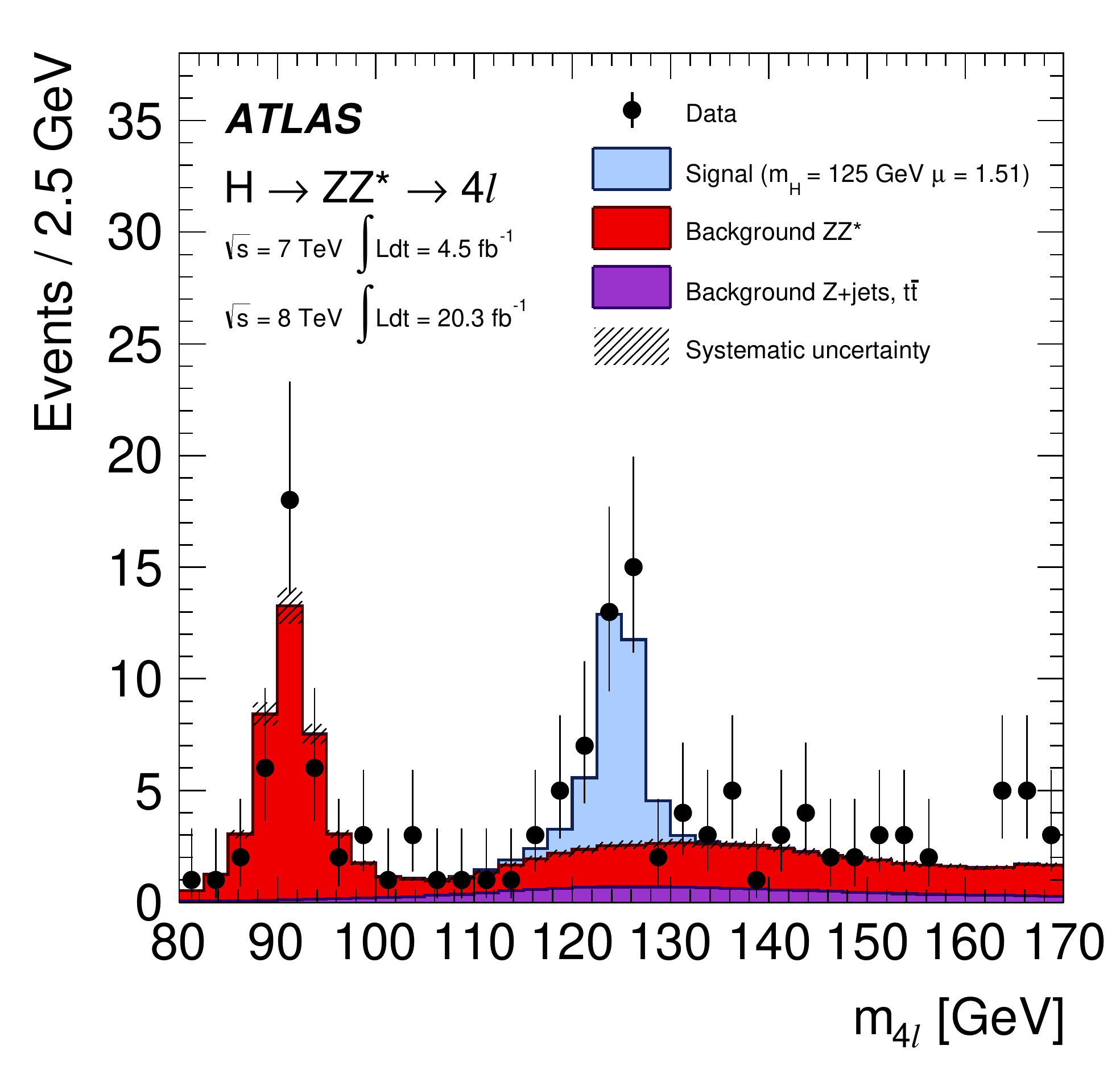}
\caption{\label{h4l_a1}The ATLAS combined $m_\mathrm{4l}$ distribution~\cite{h4l_atlas}.} 
\end{minipage}\hspace{0.04\textwidth}%
\begin{minipage}{0.47\textwidth}
\includegraphics[height=7cm]{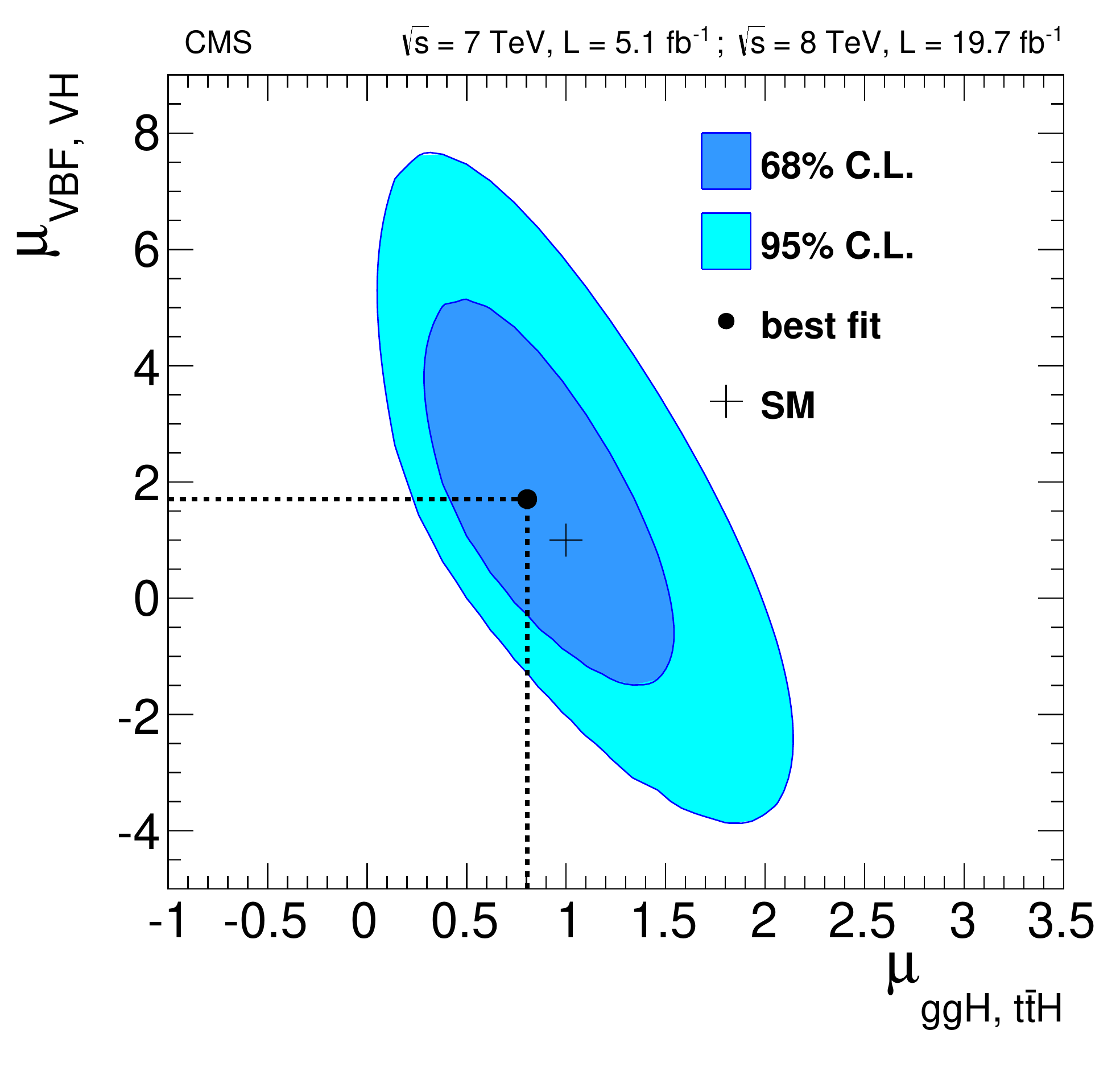}
\caption{\label{h4l_c1}The CMS measurement of the signal strength associated to ggF and $ttH$  production 
versus VBF and $VH$ production~\cite{h4l_cms}.}
\end{minipage}
\end{center}
\end{figure}
The expected rate of $H \to 4$ lepton events is low compared to the other channels presented here; however, this is compensated by the highest 
signal-over-background ratio. This channel is further characterized by a high mass resolution ($m_\mathrm{4l}$) and dominates the Higgs boson mass 
measurement together with $H \to \gamma\gamma$. The analyses~\cite{h4l_atlas,h4l_cms} first require four light leptons and are then pursued in 
subchannels depending on the flavor of the two lepton pairs. Like the $H \to \gamma\gamma$ analysis, it is split in categories targeting different production modes. 
However, the main sensitivity rests in the $0/1$ jet category aimed at gluon-gluon fusion production. The main background is continuum $ZZ^*$ production 
which is estimated from simulation and validated in control regions. The combined $m_\mathrm{4l}$ distribution for ATLAS is shown in Fig.~\ref{h4l_a1}. 
The mass peak at about 125 GeV is clearly visible.

Both the ATLAS and CMS results are in very good agreement with the SM expectation. This is illustrated in Fig.~\ref{h4l_c1} where the signal strength 
for the ggF and $ttH$ production mode is measured with respect to the VBF and $VH$ production modes. 
The observed signal significance is $6.8\sigma$ (expected: $6.7\sigma$) for
CMS and $8.1\sigma$ (expected: $6.2\sigma$) for ATLAS.

\subsection{$H \to WW$}
\begin{figure}[htb]
\begin{center}
\begin{minipage}{0.53\textwidth}
\includegraphics[height=6.0cm]{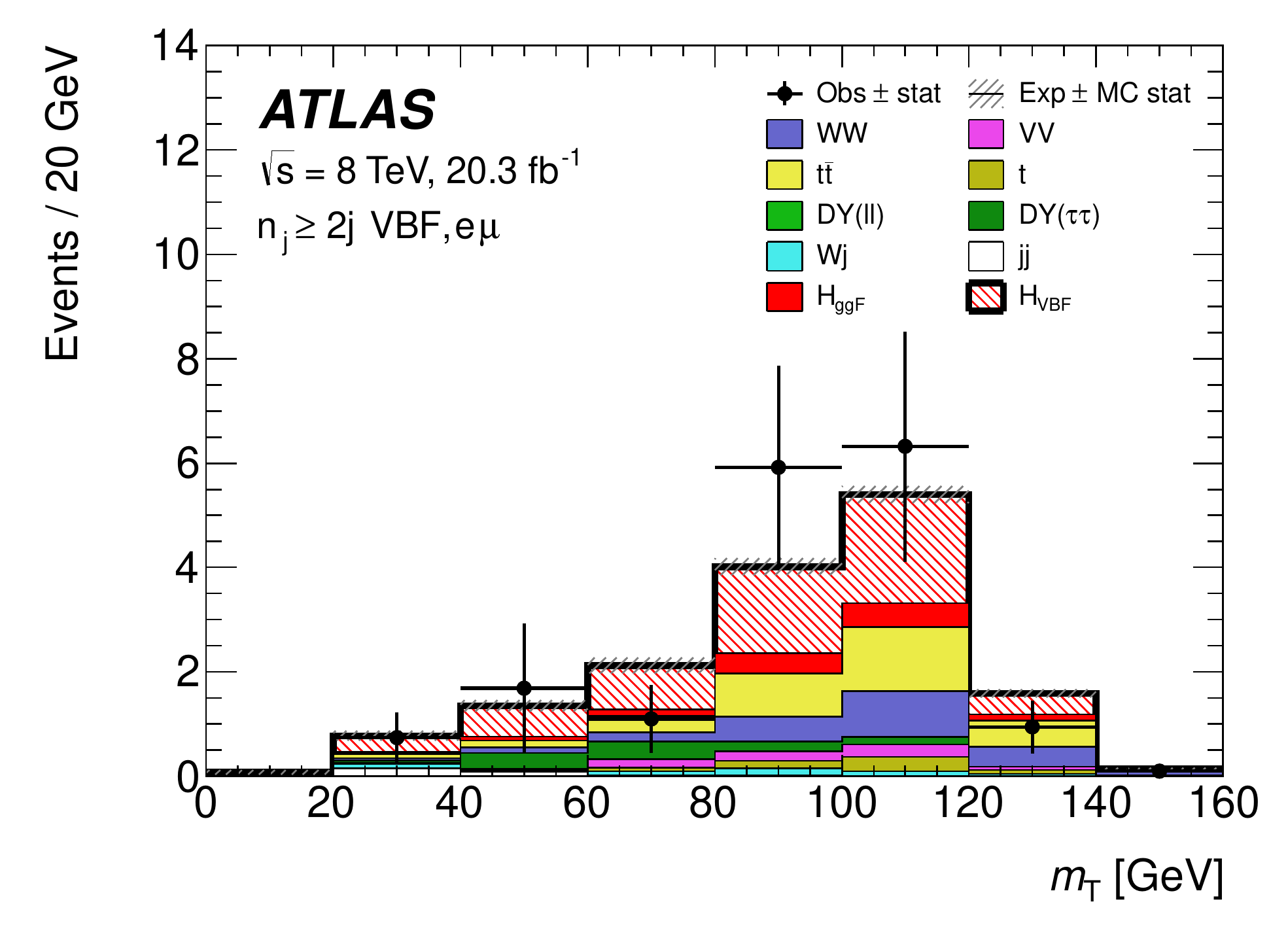}
\caption{\label{hww_a1}The ATLAS $m_T$ distribution in the category ``$e\nu \mu\nu$ + 2 (or more) jets''~\cite{hww_atlas}. 
The events are weighted by the value $\ln(1+S/B)$ of the corresponding bin of the final analysis discriminant, where 
$S$ and $B$ are the number of expected signal and background events, respectively.}   
\end{minipage}\hspace{0.04\textwidth}%
\begin{minipage}{0.41\textwidth}
\includegraphics[height=6.0cm]{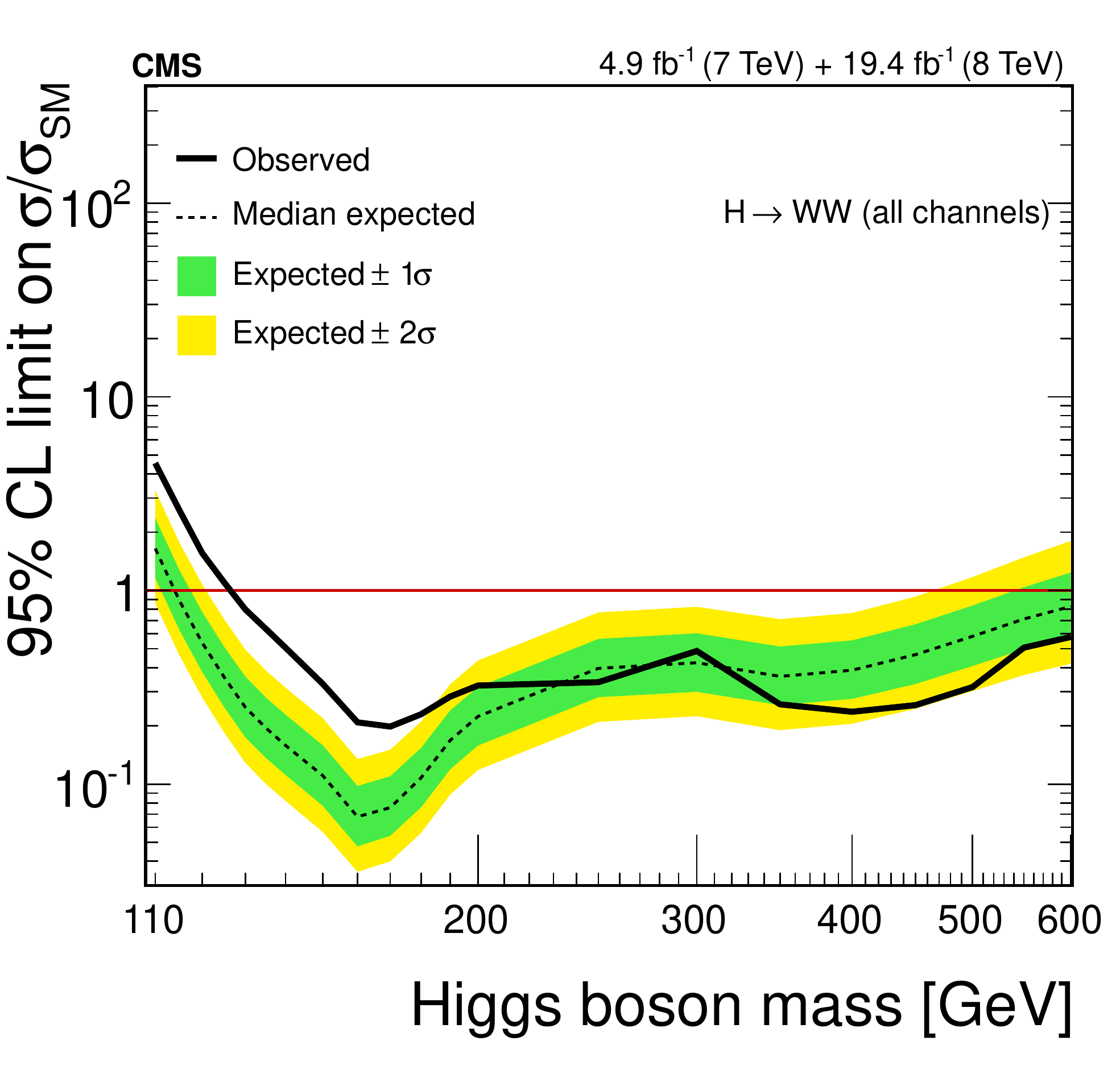}
\caption{\label{hww_c1}The CMS exclusion versus Higgs boson mass in the $WW$ channel~\cite{hww_cms}, with no assumptions on the presence of a Higgs boson. 

}
\end{minipage}
\end{center}
\end{figure}
Higgs boson decays to a $W$ boson pair offer 
a relatively high expected rate but suffer from a high irreducible $WW$ continuum background which is hard to suppress due to 
the low $m_{WW}$ resolution, caused by the presence of neutrinos in the most sensitive $W$ decay modes. The $W$+jets and top quark pair 
background is also sizable. The most sensitive subchannel features a $e\nu \mu\nu$ final state with low jet 
multiplicity~\cite{hww_atlas,hww_cms}. Requirements of additional jets or leptons aim to add extra sensitivity by exploiting 
production modes other than ggF. The $m_T$ distribution in the category ``$e\nu \mu\nu$ + 2 (or more) jets'' is shown in 
Fig.~\ref{hww_a1}. This is the $WW$ category with the highest expected signal purity.

Both the observed ATLAS and CMS rates are within about $1\sigma$ of the SM expectation. In Fig.~\ref{hww_c1}, the Higgs boson 
exclusion versus $m_H$ is shown. Visible are both the excess at about 125~GeV and the exclusion of other, heavy SM-like Higgs bosons.

\subsection{$H \to \tau\tau$}
\begin{figure}[htb]
\begin{center}
\begin{minipage}{0.47\textwidth}
\includegraphics[height=7cm]{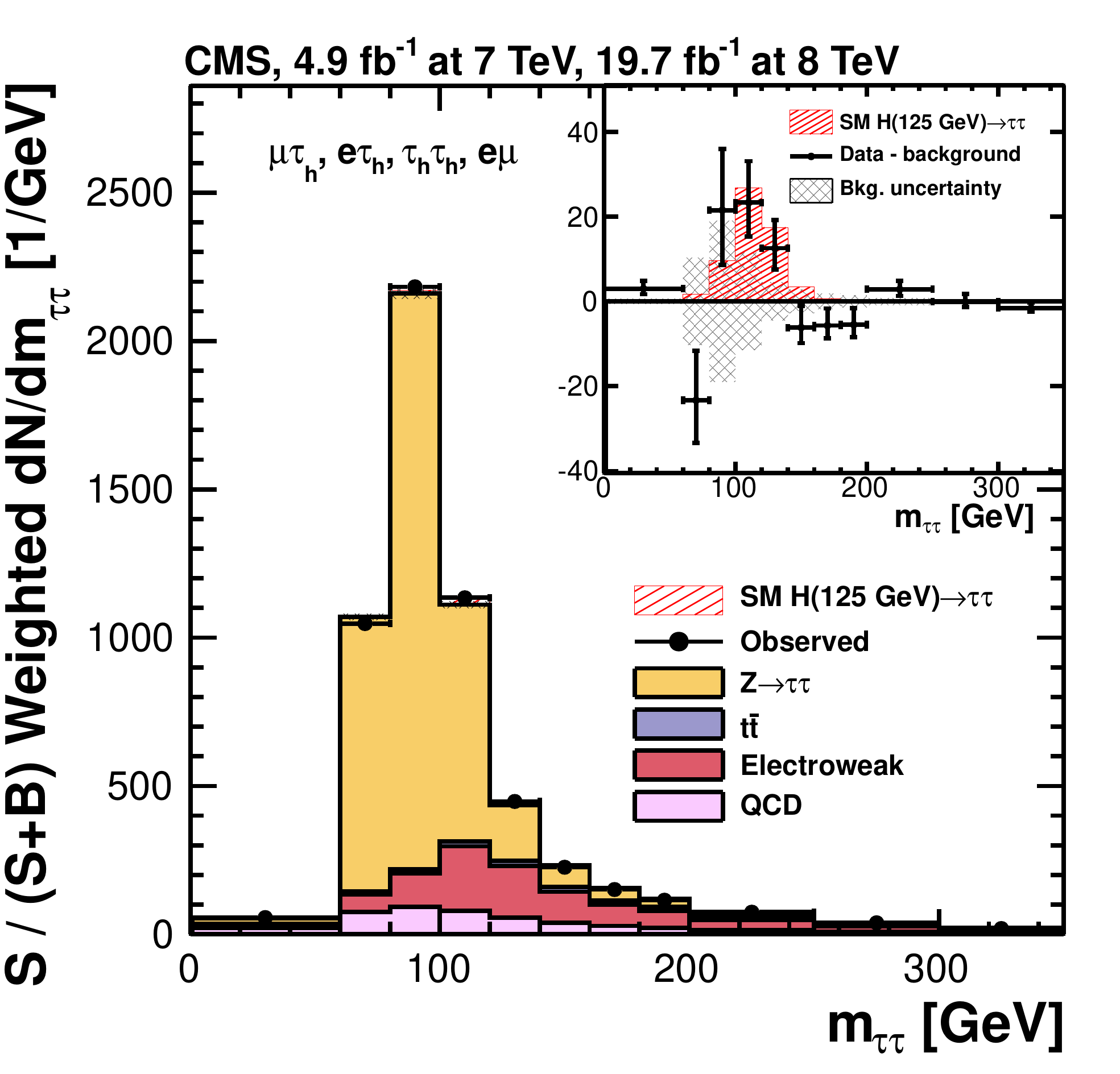}
\caption{\label{htt_c1}The $m_{\tau\tau}$ distribution estimated by CMS~\cite{htt_cms}. Events from several channels are weighted by signal purity and then combined. 
The insert at the upper right shows the background-subtracted distribution.} 
\end{minipage}\hspace{0.04\textwidth}%
\begin{minipage}{0.47\textwidth}
\includegraphics[height=7cm]{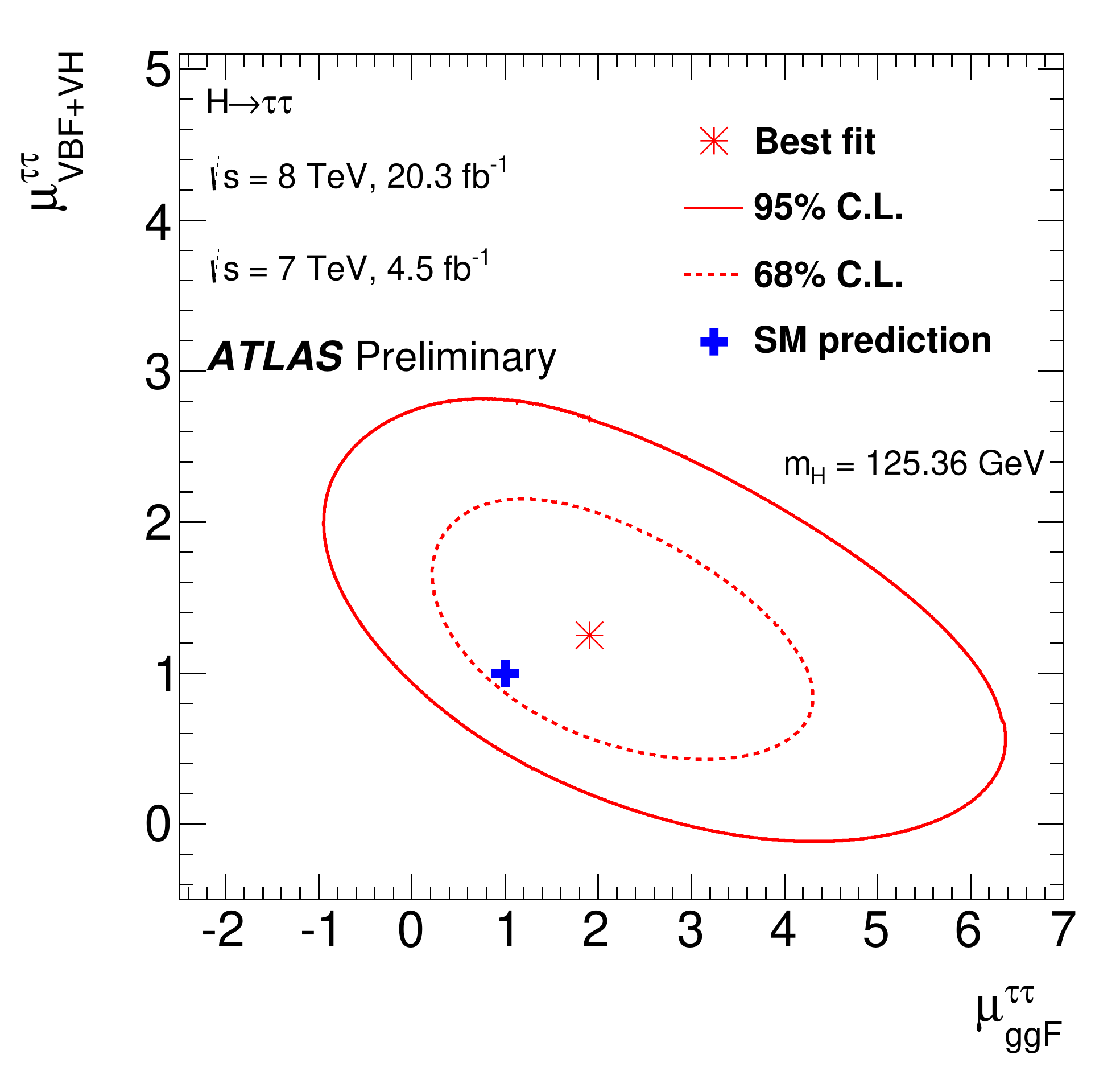}
\caption{\label{htt_a1}The ATLAS measurement of the signal strength in ggF versus VBF/$VH$~\cite{htt_atlas}. 
}
\end{minipage}
\end{center}
\end{figure}
%
The $H \to \tau\tau$ analyses are among the most complex analyses of LHC data: Six different channels, depending on 
the combination of $\tau$ lepton decays, are investigated: $ee$, $e\mu$, $\mu\mu$, $e\tau$, $\mu\tau$ and $\tau\tau$ (here, $\tau$ signifies 
a hadronic $\tau$ lepton decay, and neutrinos are omitted). For all these channels, categories motivated by the production mode can be 
established: 0/1-jet, boosted (boson candidate with high $p_T$), VBF and $VH$. CMS has also investigated $ttH$ with $H \to \tau\tau$~\cite{tth_cms}. 
The CMS analysis~\cite{htt_cms} which is based on orthogonal selection requirements implements almost one hundred categories. ATLAS~\cite{htt_atlas} uses only six; however, 
here, the analysis is based on a Boosted Decision Tree (BDT). 
Both ATLAS and CMS use 
sophisticated $m_{\tau\tau}$ estimators using kinematic information of the whole decay chain to reconstruct a mass value in spite of the presence of 
a number of undetected neutrinos. The combined $m_{\tau\tau}$ distribution is shown in Fig.~\ref{htt_c1}. The main background is due to $Z/\gamma^* \to \tau\tau$ events which 
are estimated by replacing muons in $Z \to \mu\mu$ collision data events with simulated $\tau$ leptons.

Both ATLAS and CMS report evidence for $H \to \tau\tau$ decays. ATLAS observes $4.5\sigma$ (expected: $3.5\sigma$), CMS observes $3.2\sigma$ (expected: $3.7\sigma$). This 
is the first evidence for Higgs boson decays to $\tau$ leptons. The sensitivity is mostly driven by the VBF category, as can be seen in Fig.~\ref{htt_a1}. While 
the signal strength in ggF is only weakly constrained, VBF/$VH$ decays are established at a $2\sigma$-level.

\subsection{$H \to bb$}
\begin{figure}[htb]
\begin{center}
\begin{minipage}{0.47\textwidth}
\includegraphics[height=7cm]{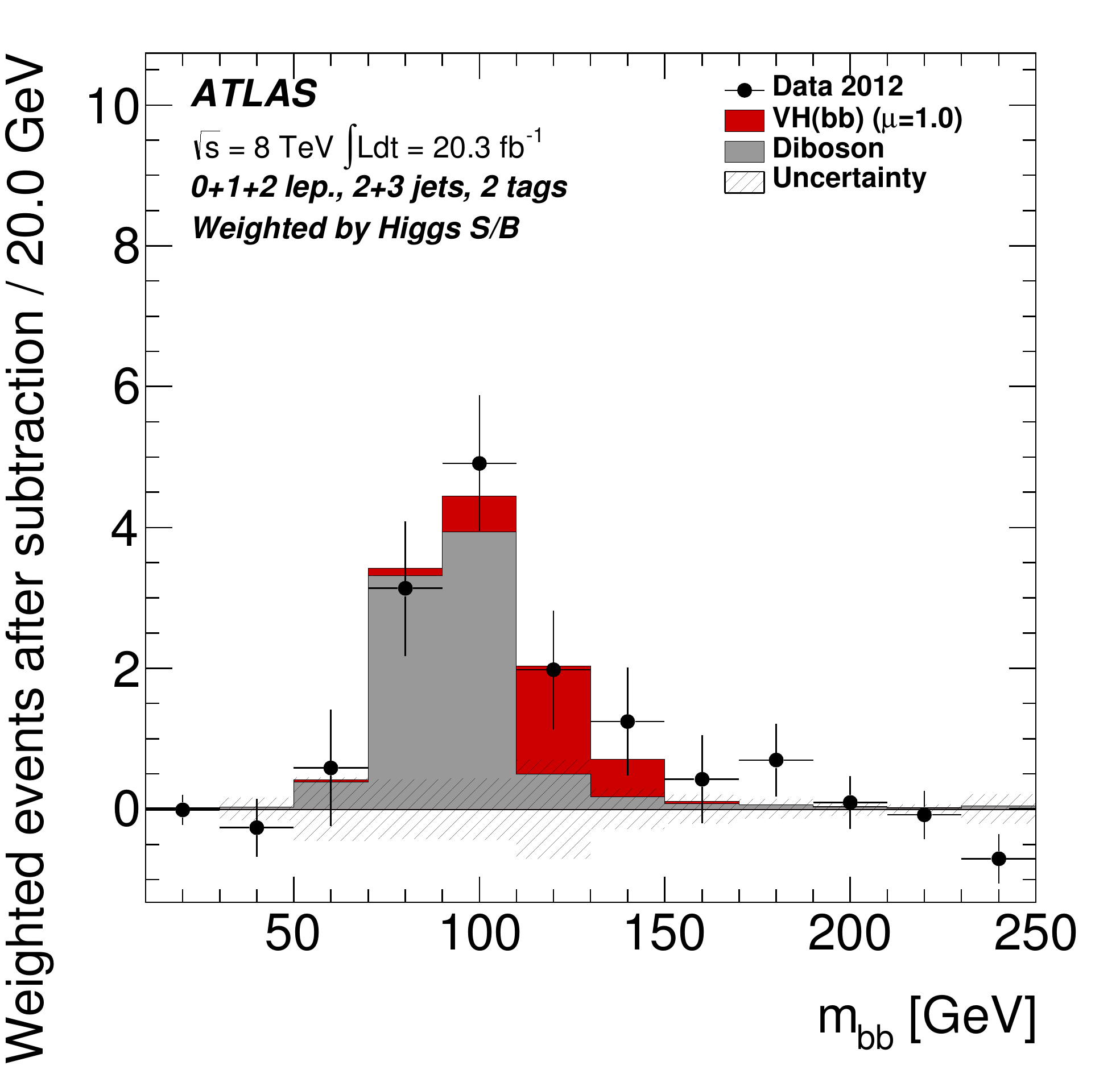}
\caption{\label{hbb_a1}The $m_{bb}$ distribution of the ATLAS analysis after subtracting all backgrounds except diboson events~\cite{hbb_atlas}.} 
\end{minipage}\hspace{0.04\textwidth}%
\begin{minipage}{0.47\textwidth}
\includegraphics[height=7cm]{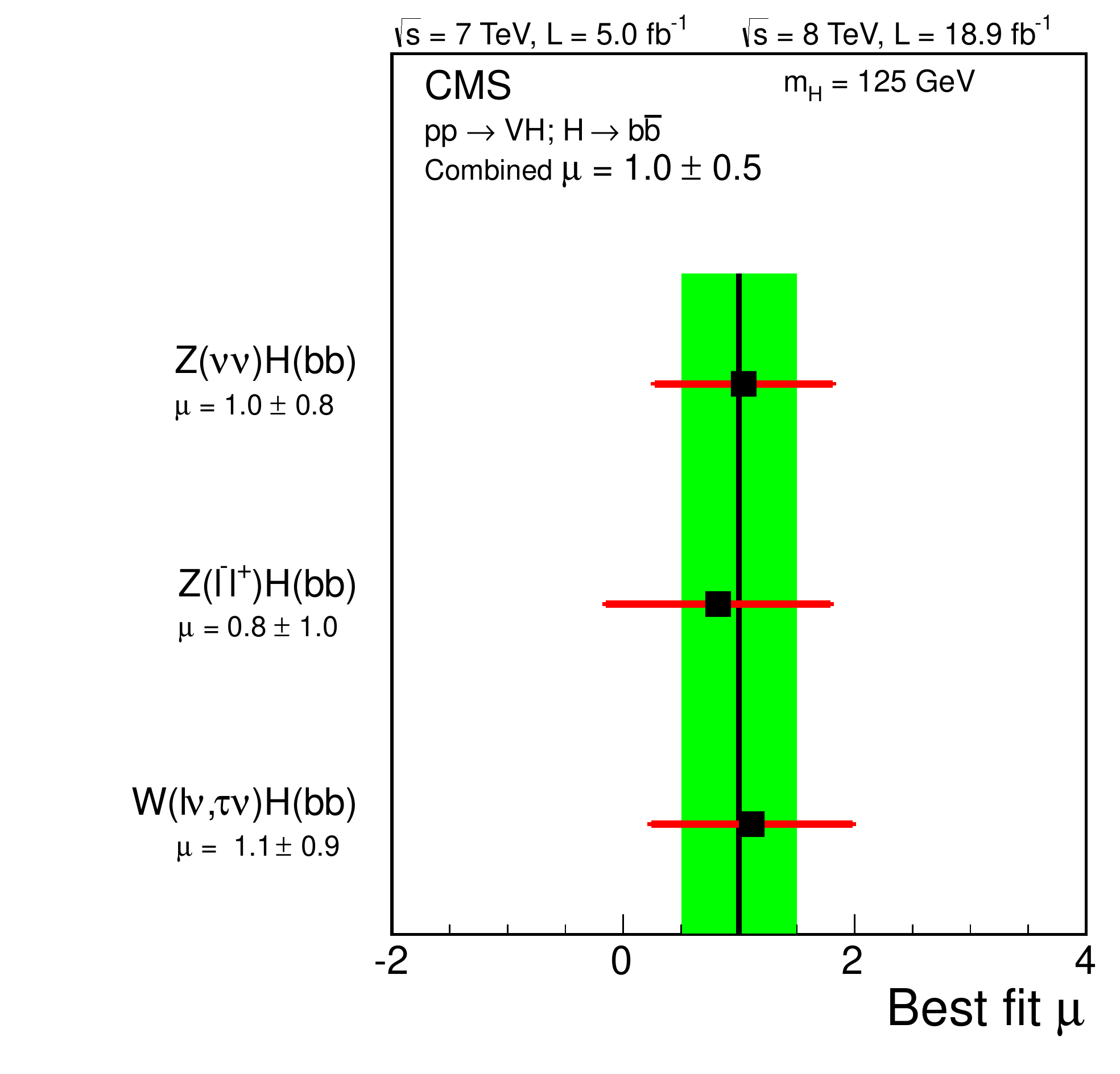}
\caption{\label{hbb_c1}The signal strength in the $H \to bb$ channel as measured by CMS, split by production modes~\cite{hbb_cms}.}
\end{minipage}
\end{center}
\end{figure}
The $H \to bb$ analysis is extremely challenging at the LHC, despite the high expected branching ratio of 
$0.58$ at $m_H=125$ GeV. The ggF production is not accessible as the signal cannot be separated from the
overwhelming non-resonant $b\bar{b}$ background.
The situation for VBF production is only marginally better. The main workhorse are thus $VH$ events; in addition, 
$ttH$ is also investigated. Both experiments use a BDT-based analysis~\cite{hbb_atlas,hbb_cms} to isolate the signal 
from the overwhelming background of $Z$, $W$, $t\bar{t}$, $VV$ and multi-jet events. The $m_{bb}$ distribution after 
background subtraction is shown in Fig.~\ref{hbb_a1}.

Both observed and expected significance by CMS are $2.1\sigma$. ATLAS observes $1.4\sigma$ while the expectation is $2.6\sigma$. The 
non-significant ATLAS deficit is driven by the $\sqrt{s}=7$~TeV data. The CMS measurement of the signal strength is shown in 
Fig.~\ref{hbb_c1}.

\subsection{$ttH$}
\begin{figure}[htb]
\begin{center}
\begin{minipage}{0.57\textwidth}
\includegraphics[height=6.3cm]{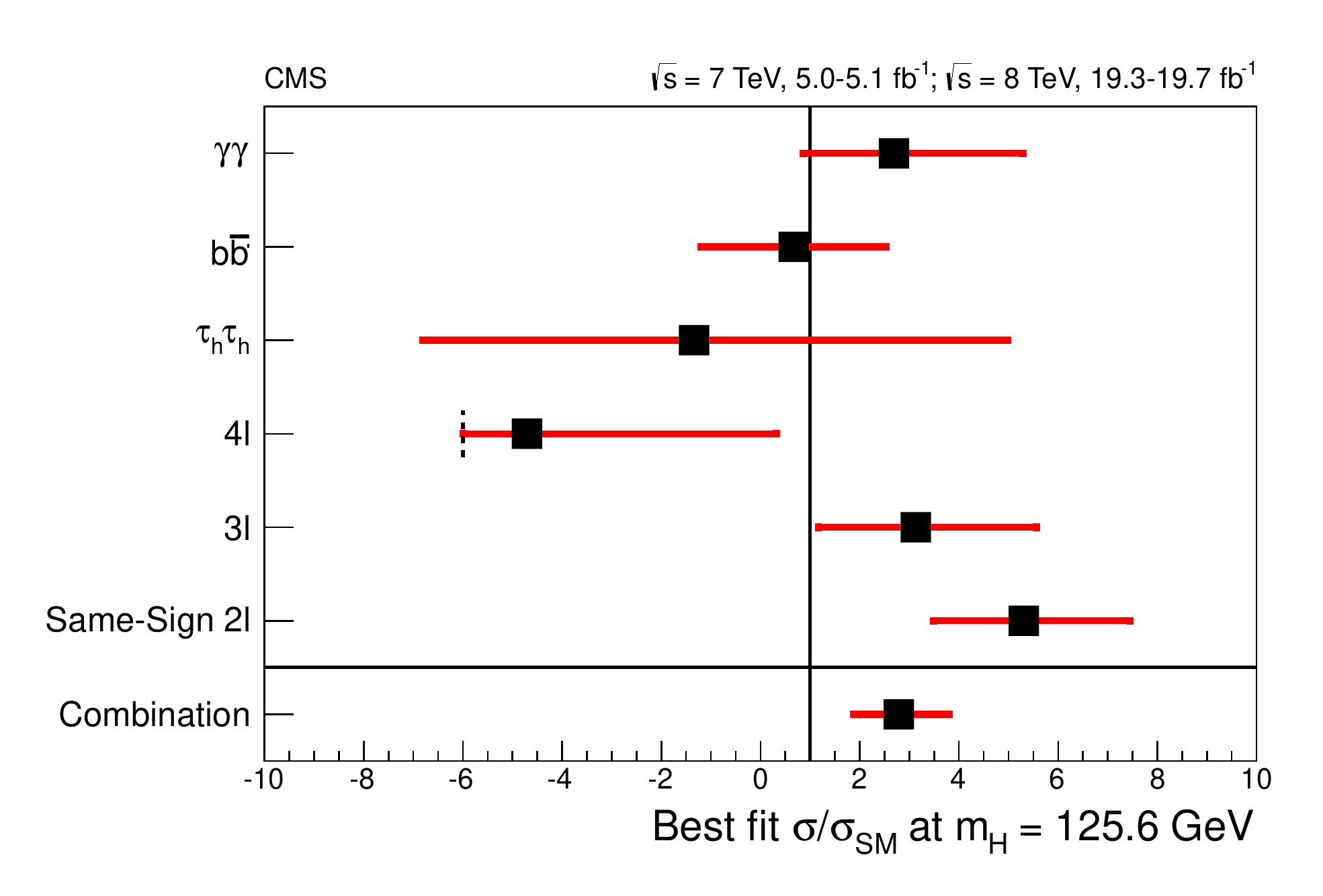}
\caption{\label{tth_c2}The measured signal strength in the CMS $ttH$ analyses~\cite{tth_cms}.}
\end{minipage}\hspace{0.04\textwidth}%
\begin{minipage}{0.37\textwidth}
\includegraphics[height=6.3cm]{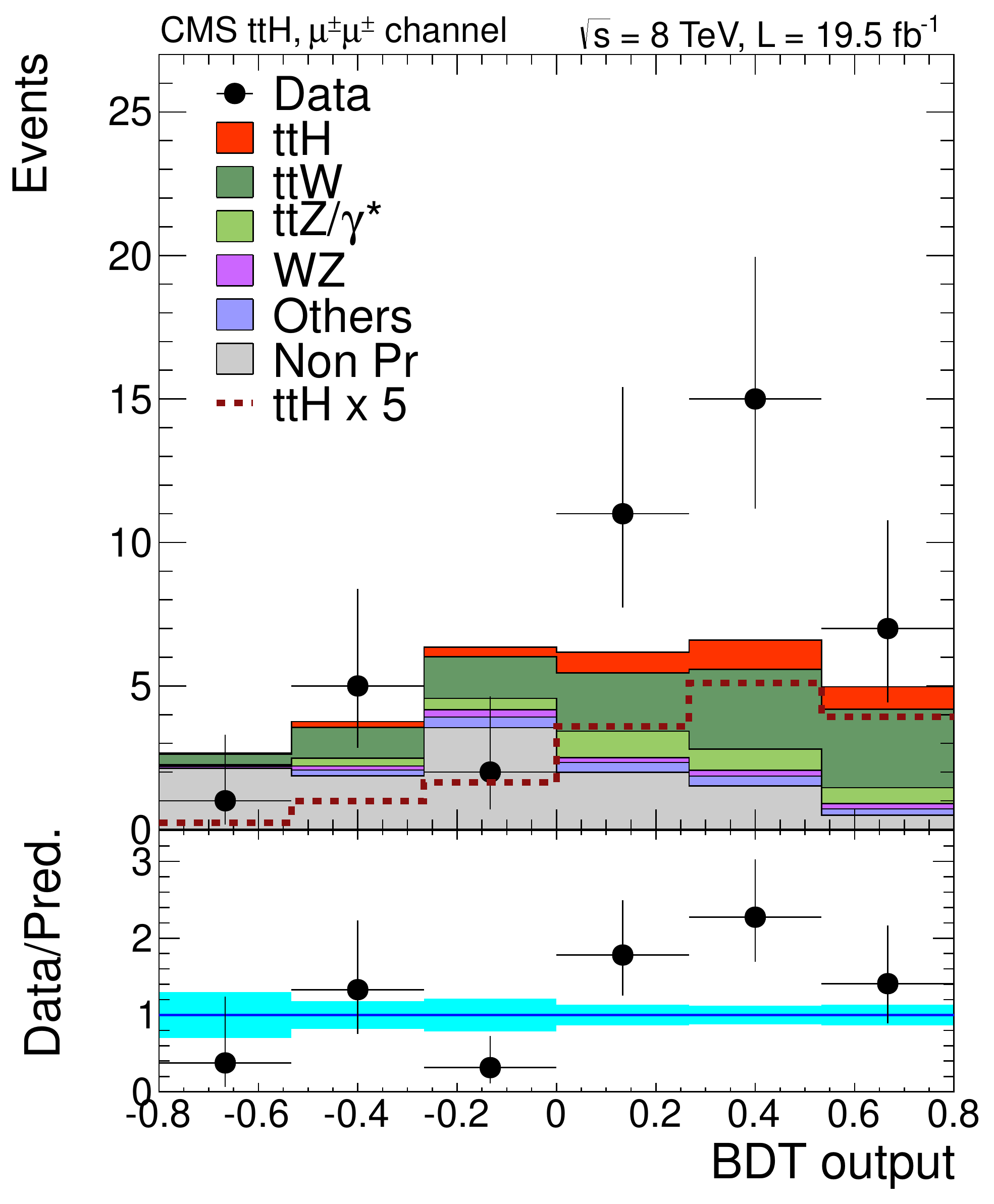}
\caption{\label{tth_c1}The BDT output for the CMS $ttH$ analysis in the same-sign two-lepton channel~\cite{tth_cms}.} 
\end{minipage}
\end{center}
\end{figure}
The $ttH$ channel poses very particular challenges and is thus in most cases not treated in the context of the usual decay mode analyses by ATLAS 
and CMS. Instead, dedicated $ttH$ analyses and combinations are produced. CMS investigates six different final states~\cite{tth_cms}, see 
Fig.~\ref{tth_c2}. The measured combined signal strength is $\mu=2.8 \pm 1.0$ which is about $2\sigma$ high compared to the SM expectation. 
The small excess is almost entirely driven by the same-sign two-lepton category. The BDT output for this category is shown in Fig.~\ref{tth_c1} 
and illustrates this point.
ATLAS investigates and combines~\cite{tth_atlas} two final states in $ttH$ production, namely $bb$ and $\gamma\gamma$. The combined ATLAS limit on $ttH$ 
production is in good agreement with the SM.

\section{Higgs boson property measurements}

The Higgs boson property measurements use the results of the analyses aiming at the various production and decay modes presented in the previous 
section. For each measurement, the channels sensitive to the property in question are combined. Among the studied properties are mass, width, 
signal strength (i.e., normalized cross section times branching ratio), coupling strength and tensor coupling structure (CP properties) of the 
Higgs boson in relation to other SM particles. In addition, differential cross sections allow further insight and additional tests of BSM scenarios.

\subsection{Mass}
\begin{figure}[htb]
\begin{center}
\begin{minipage}{0.54\textwidth}
\includegraphics[height=6.0cm]{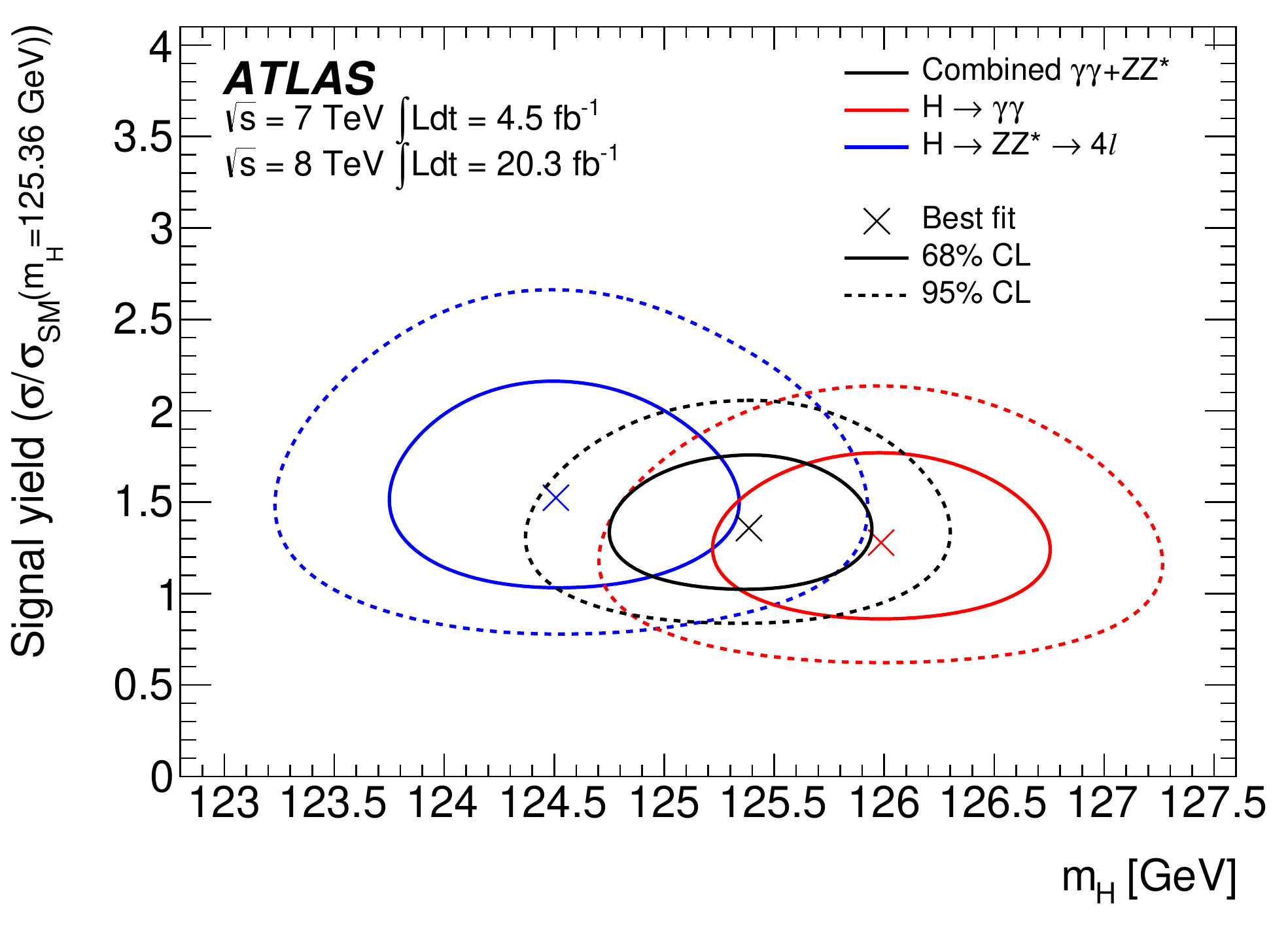}
\caption{\label{mass_a1}ATLAS $m_H$ measurement versus the signal strength evaluated at the best-fit mass~\cite{mass_atlas}.} 
\end{minipage}\hspace{0.04\textwidth}%
\begin{minipage}{0.40\textwidth}
\includegraphics[height=6.0cm]{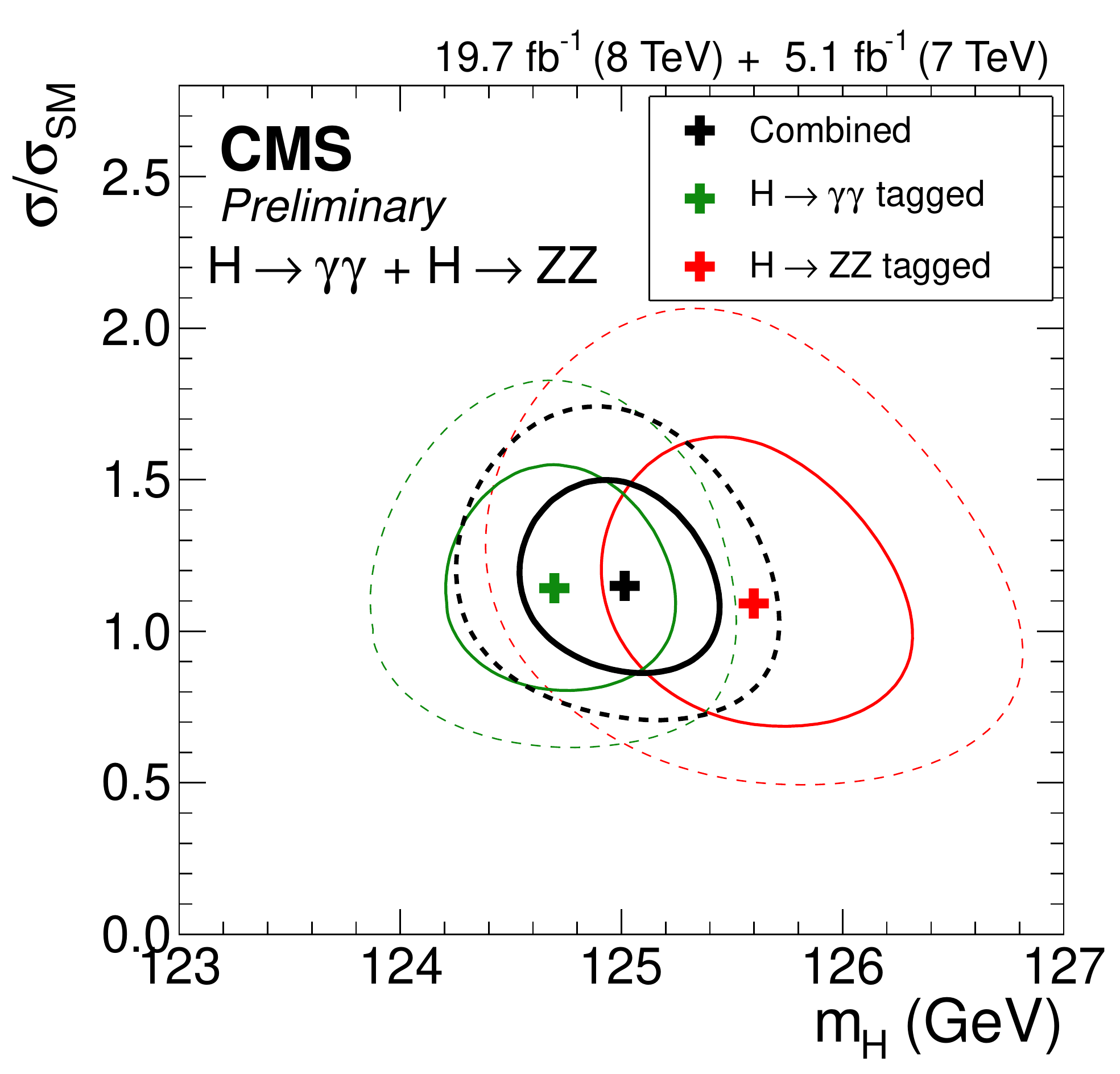}
\caption{\label{mass_c1}CMS $m_H$ measurement versus the signal strength~\cite{mass_cms}.}
\end{minipage}
\end{center}
\end{figure}
The Higgs boson mass measurement uses input from the $\gamma \gamma$ and 4-lepton decay modes. 
In both channels, the Higgs boson candidate mass can be reconstructed with high precision directly 
from its visible decay products. The ATLAS measurement~\cite{mass_atlas}, illustrated in Fig.~\ref{mass_a1}, yields 
$m_H=125.36 \pm 0.37 \mathrm{(stat)} \pm 0.18 \mathrm{(syst)}$. The CMS measurement~\cite{mass_cms}, 
see Fig.~\ref{mass_c1}, is consistent with this result: $m_H=125.03 ^{+0.26}_{-0.27} \mathrm{(stat)} ^{+0.13}_{-0.15} \mathrm{(syst)}$. 
Historically, there has been a tension between the two ATLAS measurements. However, this tension is gradually decreasing and 
currently the $\gamma\gamma$ mass is only about $2\sigma$ high with respect to the four-lepton mass. 
The two CMS mass measurements have a $1.4\sigma$ tension in the opposite direction, i.e. $m_\mathrm{4l} < m_{\gamma\gamma}$.

\subsection{Width}
\begin{figure}[htb]
\begin{center}
\begin{minipage}{0.47\textwidth}
\includegraphics[height=7cm]{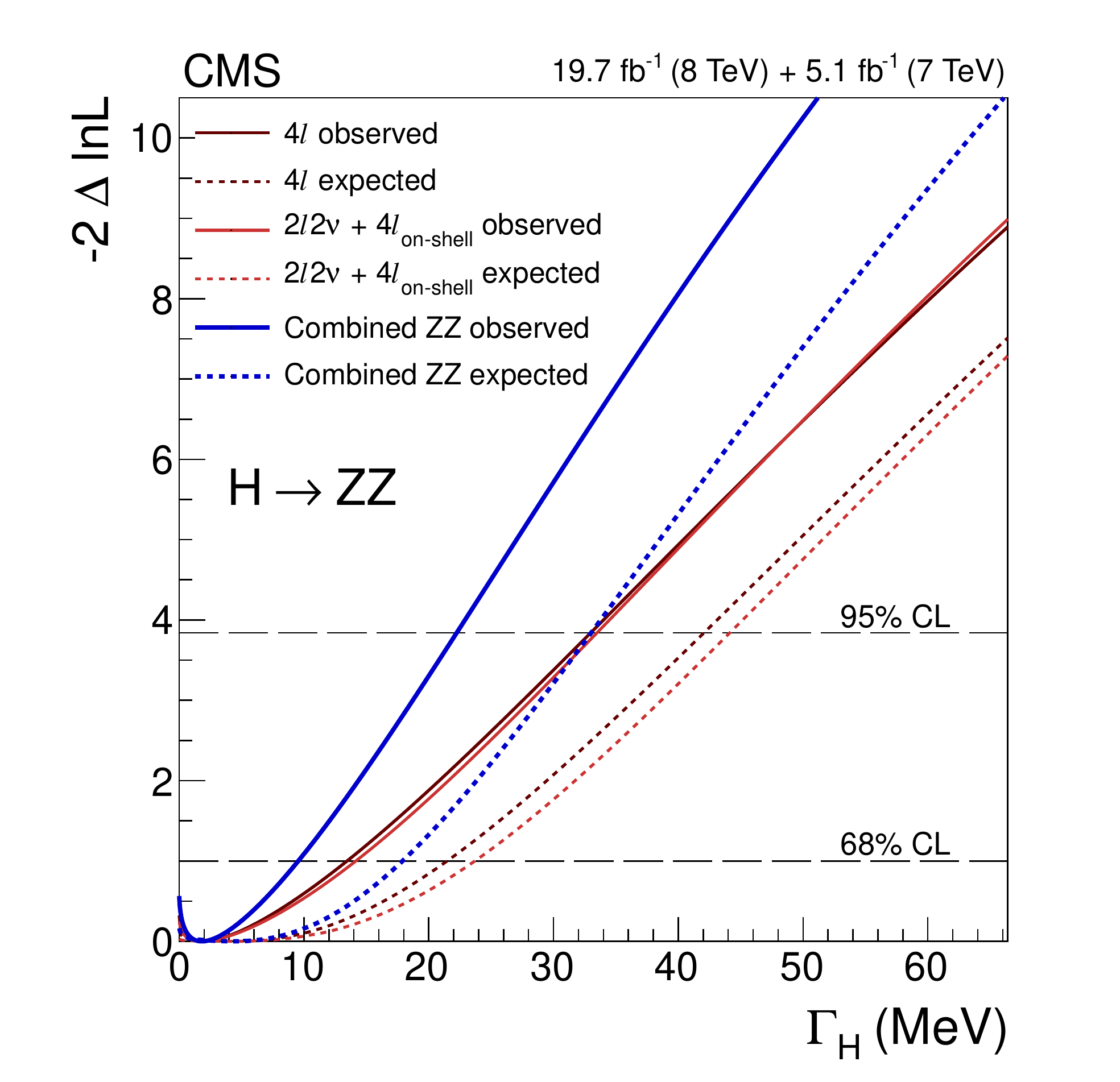}
\caption{\label{width_c1}The CMS Higgs boson width constraints from off-shell $H \to 4$ leptons signal strength measurements~\cite{width_cms}.} 
\end{minipage}\hspace{0.04\textwidth}%
\begin{minipage}{0.47\textwidth}
\includegraphics[height=7cm]{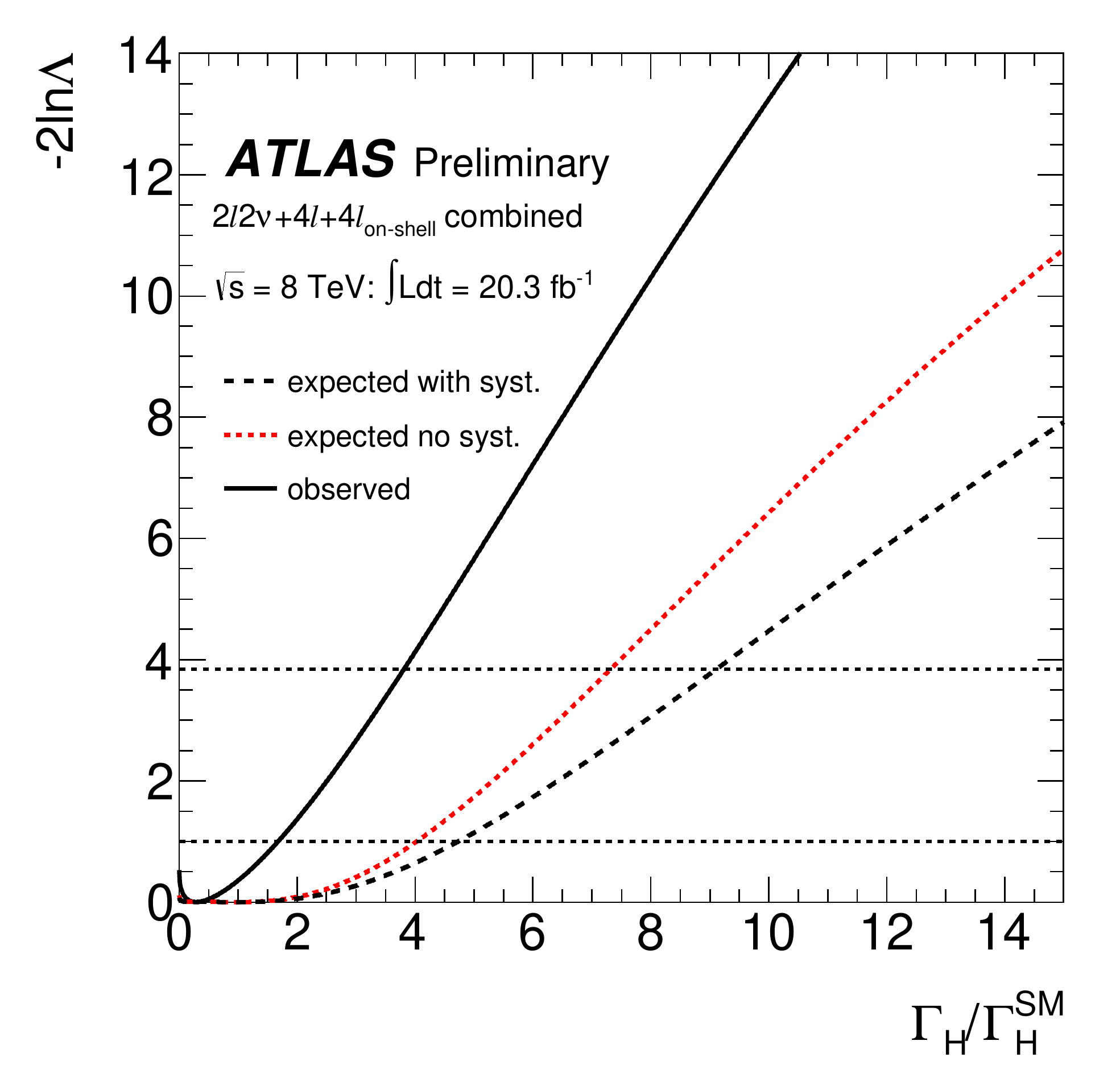}
\caption{\label{width_a1}The ATLAS Higgs boson width constraints from off-shell $H \to 4$ leptons signal strength measurements~\cite{width_atlas}.}
\end{minipage}
\end{center}
\end{figure}
The SM expectation for the Higgs boson width at $m_H=125$~GeV is about 4~MeV. There are several ways to access the Higgs boson width 
experimentally. Directly, the width can be measured by analyzing the width of the $m_\mathrm{4l}$ and $m_{\gamma\gamma}$ distributions. 
This method is limited by the experimental resolution which is about three orders of magnitude higher than the width predicted by the SM. 
The direct ATLAS width limits are 5.0 GeV ($H \to \gamma\gamma$), 2.6 GeV ($H \to 4$ leptons)~\cite{mass_atlas}; the corresponding 
CMS limits are 2.4 GeV ($H \to \gamma\gamma$)~\cite{hgg_cms} and 3.4 GeV ($H \to 4$ leptons)~\cite{h4l_cms}.

Indirect limits can be obtained by Higgs boson coupling fits (by leaving the invisible width as free fit parameter).
However, the most precise indirect measurement is based on the comparison 
of the on- and off-shell $H \to 4$ leptons signal strength. The CMS limit on the Higgs width using this technique is 
22 MeV (expected: 33 MeV), which corresponds to 5.4 (expected: 8.0) times the SM expectation~\cite{width_cms}. ATLAS uses a more flexible model 
and excludes 5-8 (expected: 7-12) times the SM expectation, depending on the ratio of the leading-order to next-to-leading order cross section
for on- and off-shell production~\cite{width_atlas}.

\subsection{Signal strength}
\begin{figure}[htb]
\begin{center}
\begin{minipage}{0.54\textwidth}
\includegraphics[height=8.0cm]{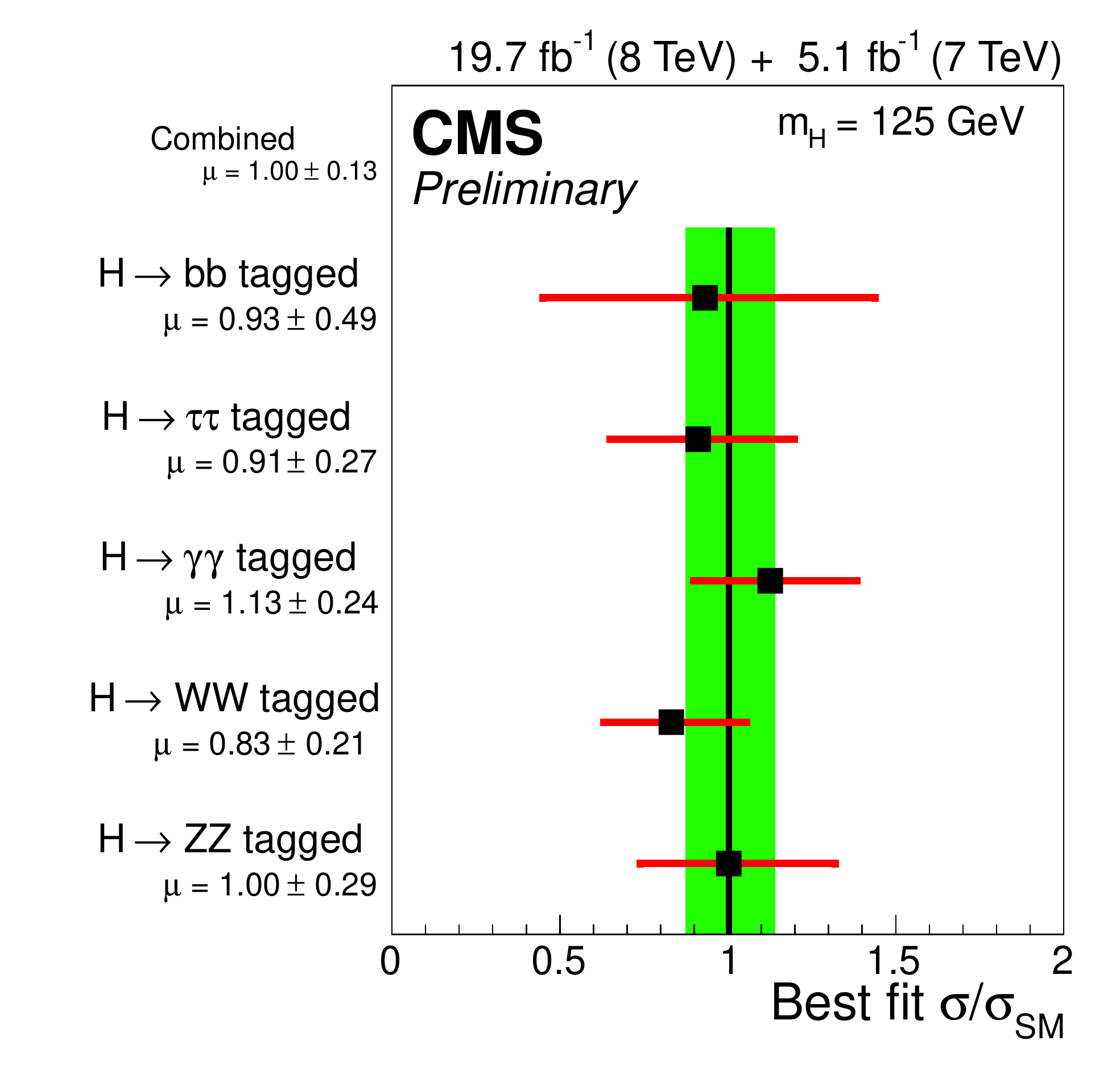}
\caption{\label{mu_c1}CMS measurement of the signal strength $\mu$~\cite{mass_cms}.}
\end{minipage}\hspace{0.04\textwidth}%
\begin{minipage}{0.4\textwidth}
\includegraphics[height=7.5cm]{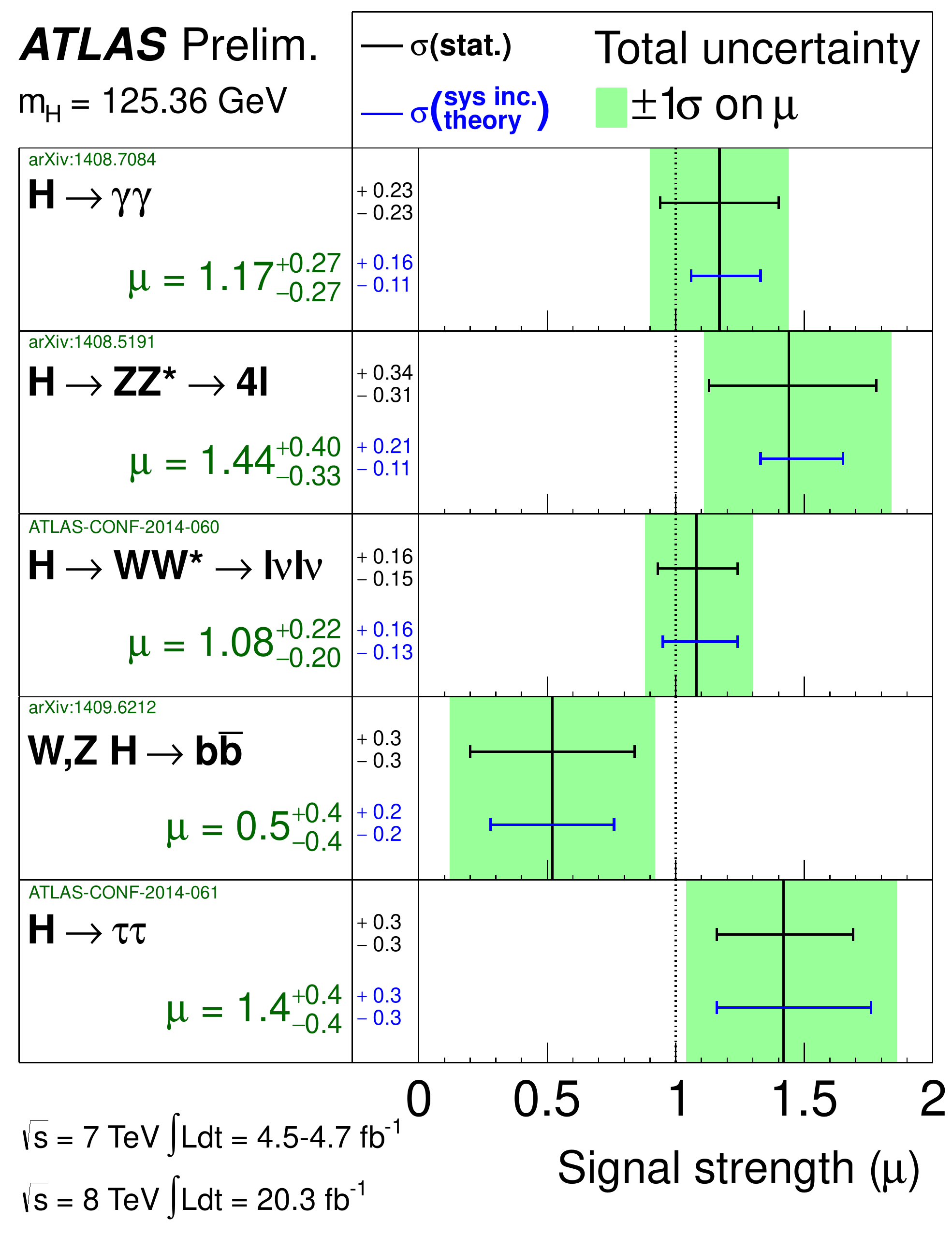}
\vspace{0.5cm}
\caption{\label{mu_a1}ATLAS measurement of the signal strength $\mu$~\cite{mu_atlas}.} 
\end{minipage}
\end{center}
\end{figure}
The signal strength $\mu$, defined as measured cross section times branching ratio of a given process divided 
by the SM expectation, is an important test for the validity of the SM. With the data accumulated in the LHC Run 1, 
no significant deviation is observed. The combined CMS result is $\mu = 1.00 \pm 0.13$, and the results for each decay 
channel are consistent with the SM within $1\sigma$~\cite{mass_cms}, see Fig.~\ref{mu_c1}. 
The ATLAS result is slightly high, $\mu = 1.30 \pm 0.18$. Note that this combined value has not been updated after the release 
of the most recent results which are summarized in Fig.~\ref{mu_a1}. The ATLAS results for all channels agree with the SM 
expectation within about $1\sigma$~\cite{mu_atlas}. 

\subsection{Coupling strength}
\begin{figure}[htb]
\begin{center}
\begin{minipage}{0.57\textwidth}
\includegraphics[height=6.0cm]{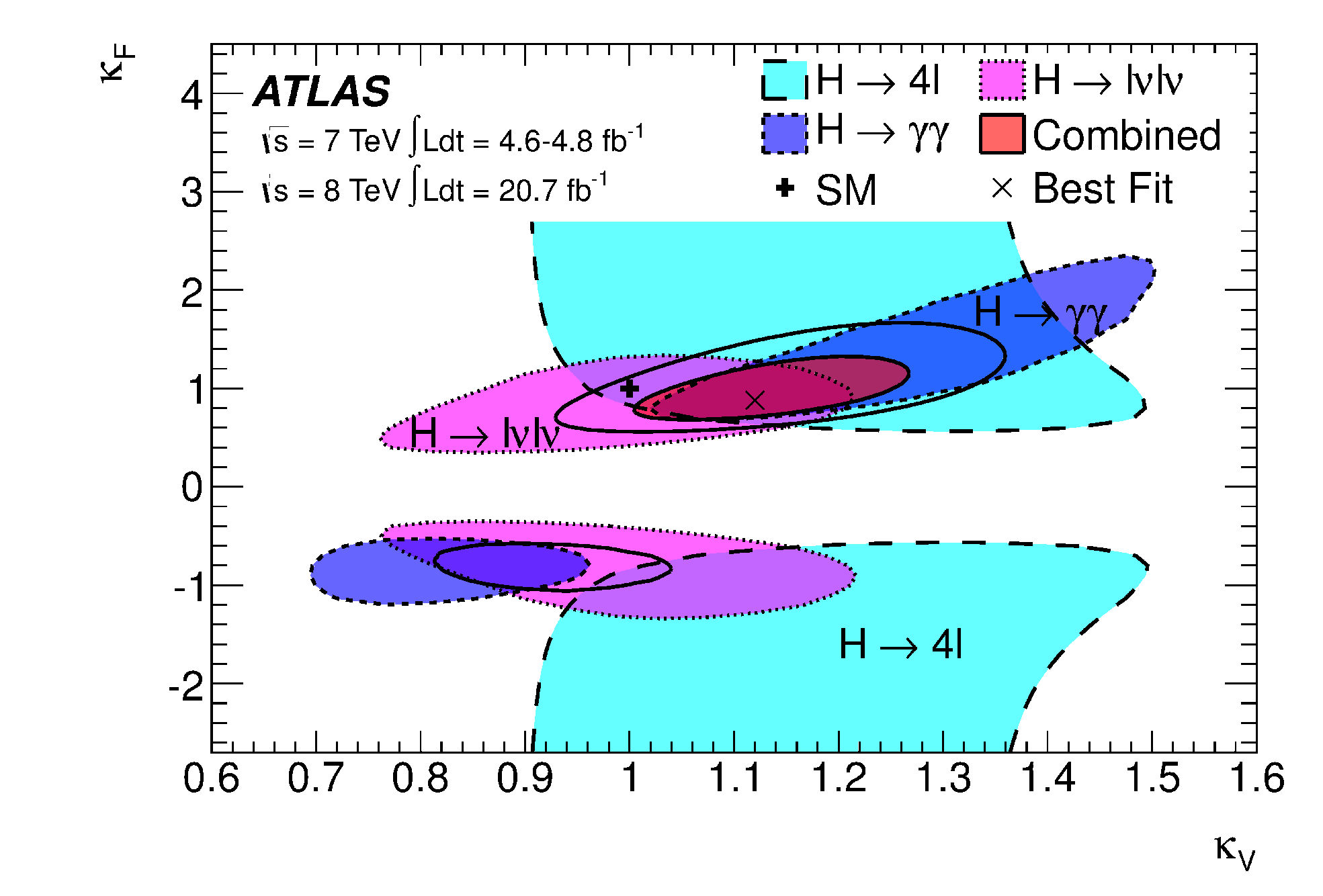}
\caption{\label{kappa_a1}ATLAS measurement of $\kappa_V$ versus $\kappa_F$~\cite{kappa_atlas}.}
\end{minipage}\hspace{0.04\textwidth}%
\begin{minipage}{0.37\textwidth}
\includegraphics[height=6.0cm]{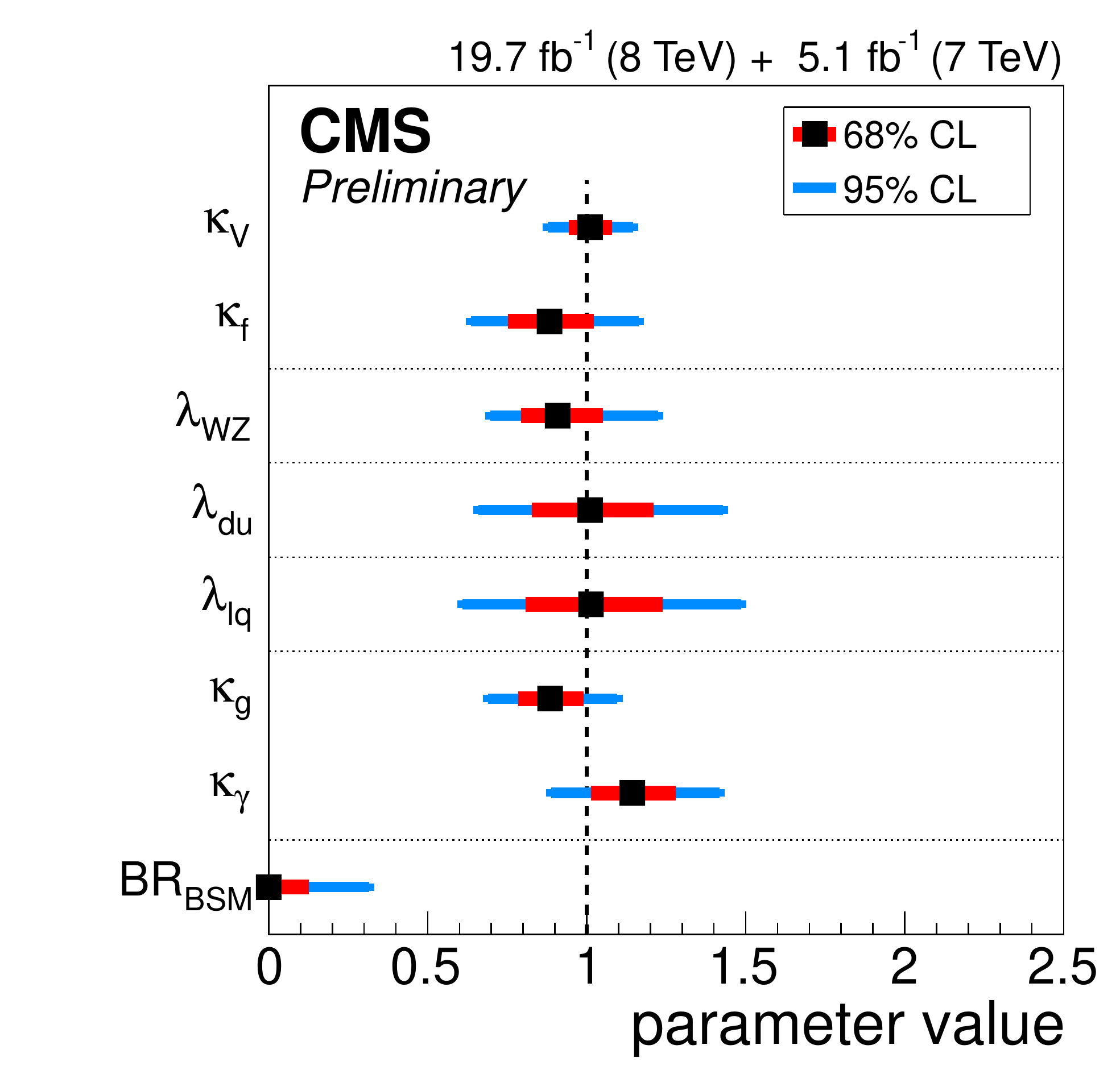}
\caption{\label{kappa_c1}CMS measurement of coupling strength parameters~\cite{mass_cms}.} 
\end{minipage}
\end{center}
\end{figure}
The coupling strength factors $\kappa_i$ are a leading-order-inspired parametrisation of the Higgs boson coupling to a particle 
or particle class $i$ with respect to the SM expectation. $\lambda_{jk}$ is used to denote the ratio of two values $\kappa_j$ and 
$\kappa_k$. The measurements are typically a result of a global fit of a large subset of the Higgs boson analyses and due to the 
large number of free parameters, some assumptions have to be made to obtain sensible results with the data of the LHC run 1. 
Usual assumptions are those on the Higgs width (to be SM-like) or of a universal coupling strength scaling, e.g. of all fermions or all up-type quarks.
An important SM test is the scaling of fermion couplings versus boson couplings, shown in Fig.~\ref{kappa_a1}. The result is consistent 
with the SM expectation~\cite{kappa_atlas}. Further tests involve the custodial symmetry ($W$ versus $Z$ coupling), up- versus down-type quark couplings 
or lepton versus quark couplings. As shown in Fig.~\ref{kappa_c1}, all results agree with the SM prediction~\cite{mass_cms}.

\subsection{Tensor coupling structure}
\begin{figure}[htb]
\begin{center}
\includegraphics[height=4.7cm]{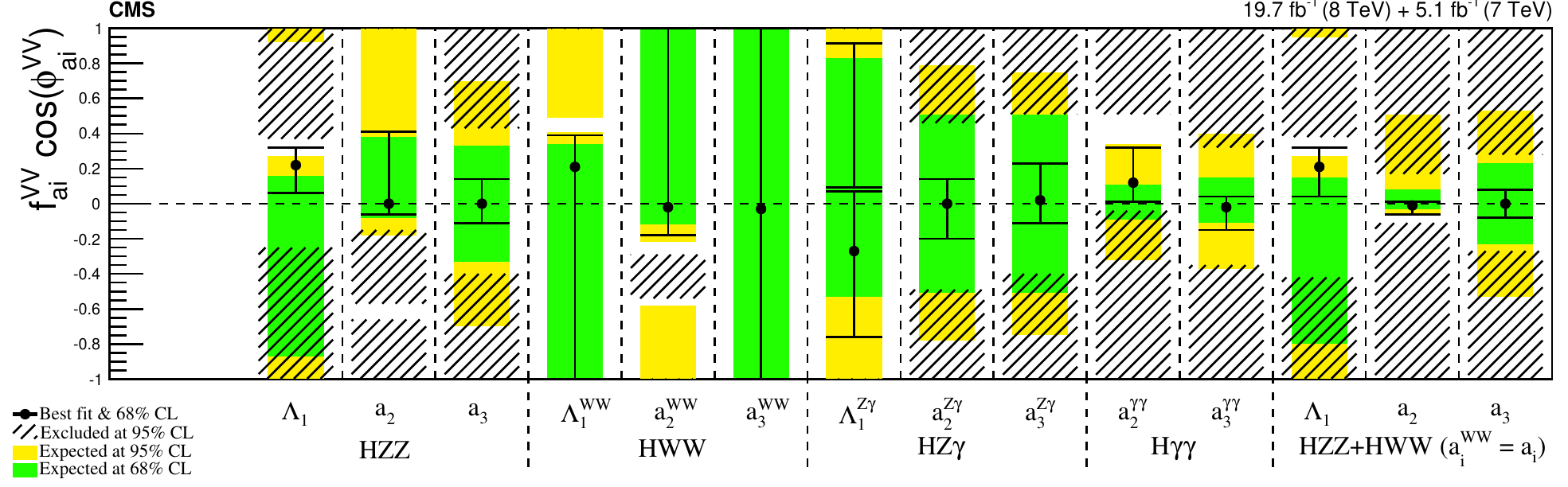}
\caption{\label{cp_c1}CMS test for anomalous components of the tensor coupling structure of Higgs bosons 
to various other particles~\cite{cp_cms}.}
\end{center}
\end{figure}
The study of the spin and CP properties of the discovered boson are essential for the claim of the discovery of the first fundamental scalar. 
To this date, all measurements are in agreement with the SM but since the parameter space of alternative models is continuous in several 
dimensions, no generic exclusion of e.g. spin-2 models has been possible. However, a large set of the best-motivated models has been tested 
with the result that the SM hypothesis is favored~\cite{cp_cms,cp_atlas}. This is illustrated in Fig.~\ref{cp_c1} where measurements of several 
parameters for the spin-0 case are made and all are found to be consistent with the SM expectation. Today, all alternatives to a spin-0 boson 
are disfavored and large anomalous contributions to the CP structure of the Higgs boson couplings are excluded. However, small or medium-sized 
CP-odd or anomalous CP-even admixtures are still feasible.

\subsection{Differential cross sections}
\begin{figure}[htb]
\begin{center}
\begin{minipage}{0.47\textwidth}
\includegraphics[height=7cm]{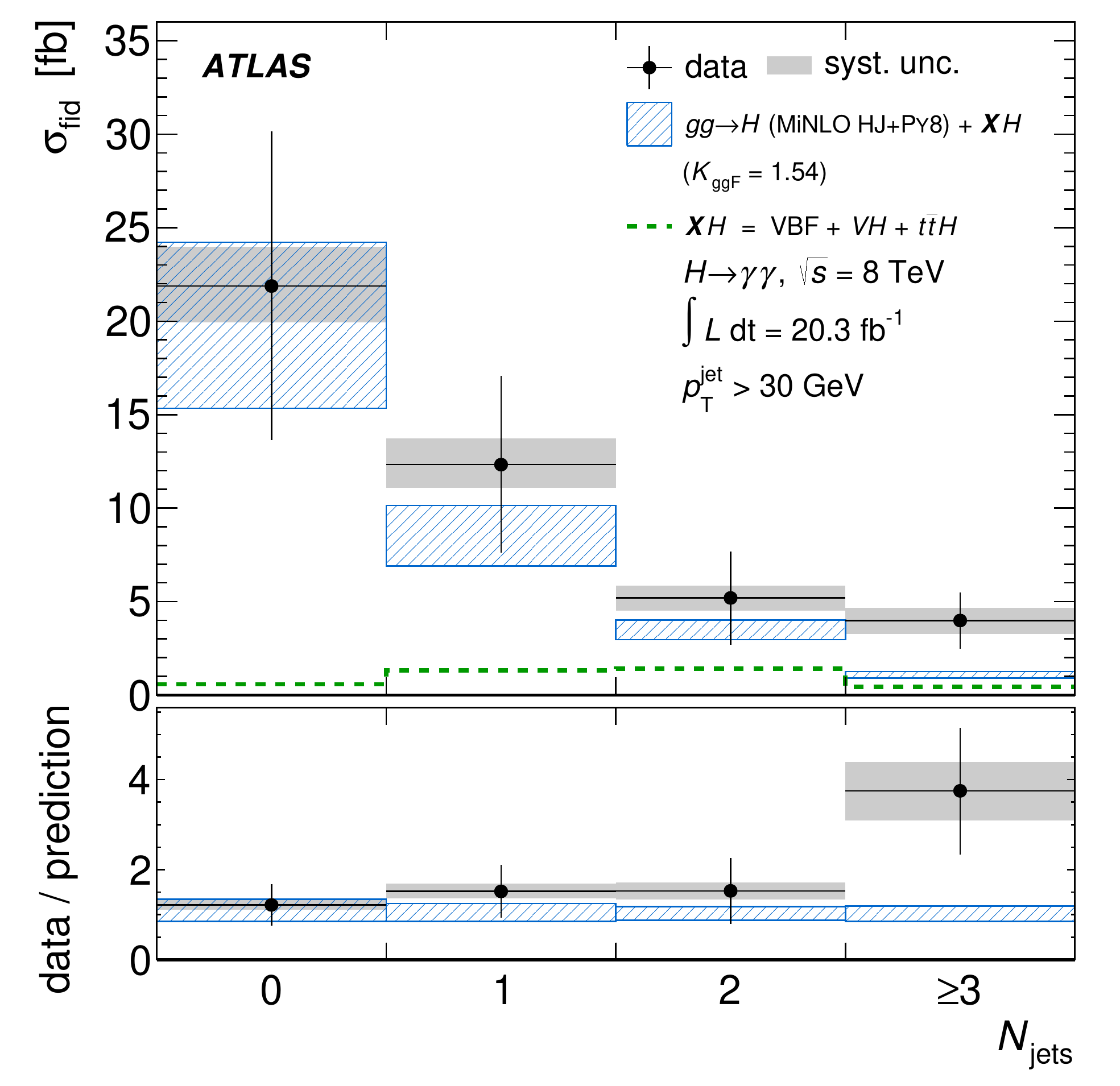}
\caption{\label{diff_a2}The jet multiplicity in ATLAS $H \to \gamma\gamma$ events~\cite{diff_gg_atlas}.}
\end{minipage}\hspace{0.04\textwidth}%
\begin{minipage}{0.47\textwidth}
\includegraphics[height=7cm]{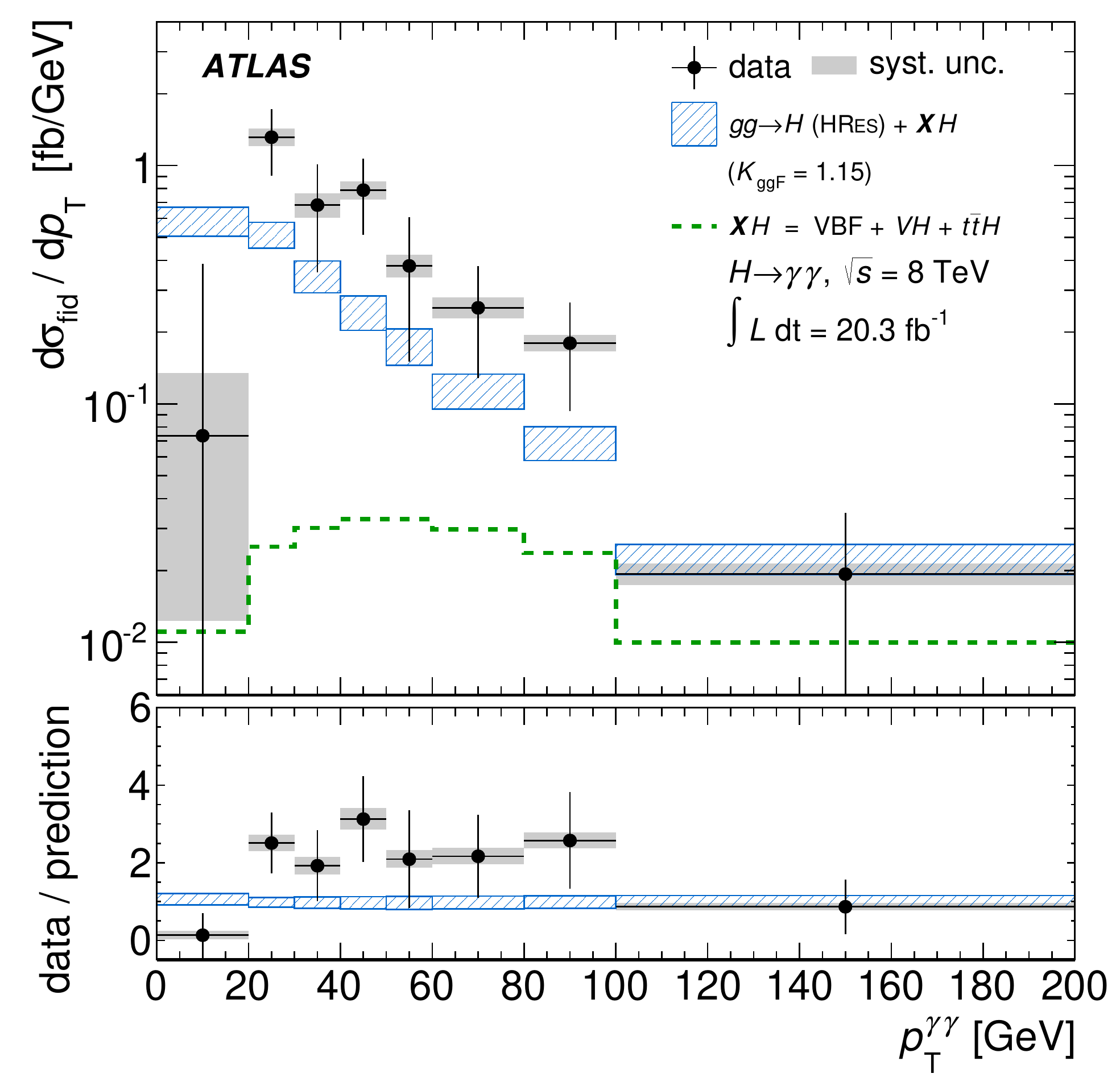}
\caption{\label{diff_a1}The $p_T$ distribution of the Higgs boson candidate in ATLAS $H \to \gamma\gamma$ events~\cite{diff_gg_atlas}.} 
\end{minipage}
\end{center}
\end{figure}
By the end of LHC Run 1, sufficient data have been collected to allow for the first 
publications of differential distributions in Higgs boson events, namely in ATLAS 
studies of the $H \to \gamma\gamma$~\cite{diff_gg_atlas} and $H \to 4$ leptons~\cite{diff_4l_atlas} 
channels. The resulting contributions are compared to predictions of the SM, but model builders can 
compare them to predictions of any competing models. Some of the aspects which are probed 
are the jet multiplicity in Higgs boson events (to test the QCD structure of SM calculations and 
models), see Fig.~\ref{diff_a2}, the transverse momentum of the Higgs boson candidate (to test for BSM effects hidden in 
loops, e.g. involving top quarks), see Fig.~\ref{diff_a1}, or the difference of the azimuthal angle 
of two of the jets in the event (to test for effects of an anomalous tensor coupling structure). 
The results, within large statistical uncertainties, are broadly in 
line with the SM predictions.

\section{BSM Higgs boson searches}
The programme of BSM Higgs boson searches at the LHC is very extensive and cannot be 
presented exhaustively here. In the following, highlights of searches for Higgs bosons 
predicted by the Minimal Supersymmetric extension of the SM (MSSM)
are presented, and other BSM Higgs boson searches are listed and their conclusions are 
summarized.

\subsection{MSSM Higgs boson searches}
\begin{figure}[htb]
\begin{center}
\begin{minipage}{0.40\textwidth}
\includegraphics[height=6.1cm]{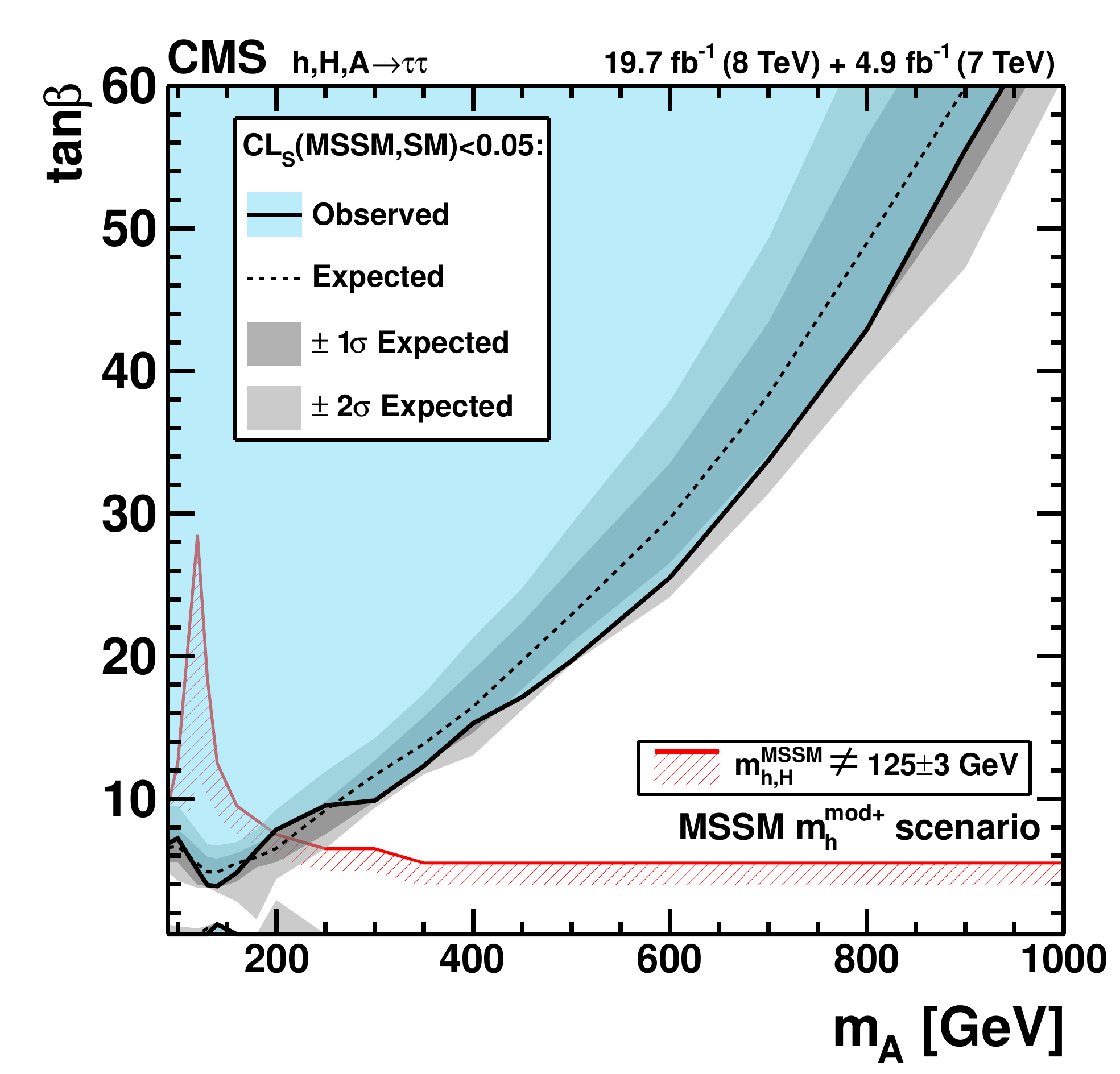}
\caption{\label{mssm_c1}The CMS $m_A$-$\tan\beta$ exclusion contour from neutral MSSM Higgs boson searches~\cite{mssmn_cms}, 
for the $m_h^{\mathrm{mod+}}$ scenario.} 
\end{minipage}\hspace{0.04\textwidth}%
\begin{minipage}{0.54\textwidth}
\includegraphics[height=6.1cm]{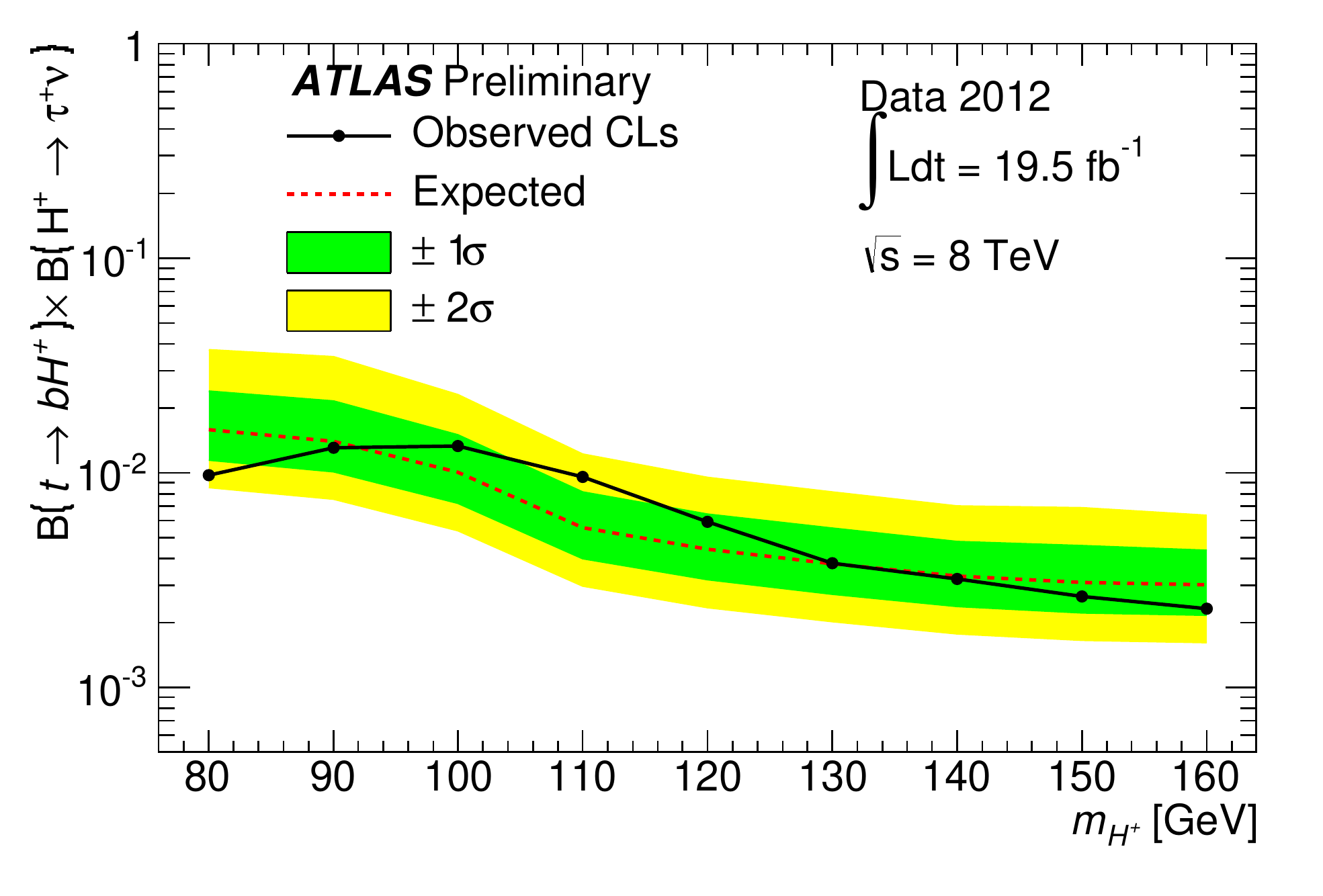}
\caption{\label{mssm_a1}ATLAS upper limit on the branching ratio BR($t \to bH^+$)~\cite{mssmc_atlas}.}
\end{minipage}
\end{center}
\end{figure}
The MSSM predicts the existence of five Higgs bosons, three of them neutral 
and a charged pair. The most sensitive search for neutral MSSM Higgs bosons 
for most of the parameter space is via their $\tau\tau$ decays. The current CMS limit~\cite{mssmn_cms} 
in the $m_h^{\mathrm{mod+}}$ scenario~\cite{mssm_benchmark} is shown in Fig.~\ref{mssm_c1}. 
Due to the charged Higgs boson search result, the option to identify the discovered Higgs boson 
(at $m_H=125$) GeV with the heavy CP-even neutral MSSM Higgs boson has been strongly disfavored. 
Presently, BR($t \to bH^+$) above about 1\% are excluded for $m_{H^+} < 160$ GeV. For charged Higgs bosons 
heavier than the top quark, a sizable region at moderate and high $\tan\beta$ has been excluded~\cite{mssmc_atlas}.

\subsection{Other BSM Higgs boson searches}
There is a vast number of BSM Higgs boson searches with ATLAS and CMS in addition to the $A/H \to \tau\tau$ and $H+ \to \tau\nu$ 
searches introduced in the previous section. A non-exhaustive list of these searches is given here:
%
\begin{itemize}

\item Other (N)MSSM-inspired searches
\begin{itemize}
\item $H^+ \to c\bar{s}$~\cite{cs_atlas,cs_cms}
\item $H^+ \to t\bar{b}$~\cite{tb_cms}
\item $H/A \to \mu\mu$~\cite{mssm_mumu_atlas,mssm_mumu_cms}
\item $H/A \to b\bar{b}$~\cite{mssm_bb_cms}
\item $H/A \to WW$~\cite{2hdm_ww_atlas}
\item $a_1 \to \gamma\gamma$~\cite{agg_atlas}
\item $a_1 \to \mu\mu$~\cite{amm_atlas,amm_cms}
\end{itemize}

\item Generic Higgs boson searches
\begin{itemize}
\item Heavy Higgs, $H \to WW / ZZ$~\cite{heavy_atlas,heavy_cms}
\item Invisible Higgs, $ZH$~\cite{inv_atlas,inv_cms}
\item Doubly charged Higgs, $H^{++}$~\cite{hpp_atlas,hpp_cms}
\item Lepton flavor violation, $H \to \tau\mu$~\cite{lfv_cms}
\item Flavor-changing neutral current, $t \to cH$~\cite{fcnc_atlas,fcnc_cms}
\item Fermiophobic Higgs~\cite{fp_atlas,fp_cms}
\item Higgs in 4$^\mathrm{th}$-generation models~\cite{4g_atlas,fp_cms}
\end{itemize}

\item Indirect search via Higgs boson property measurements~\cite{bsm_coup_atlas,mass_cms}

\item Higgs-to-Higgs decays, Higgs pair production
\begin{itemize}
\item $HH$ or $X \to HH$~\cite{pair_atlas,pair_atlas2,pair_cms,pair2_cms}
\item $H \to aa$~\cite{aa_cms}
\item $A \to ZH$~\cite{azh_cms}
\item $H \to H^+ W$ with $H^+ \to Wh$~\cite{casc_atlas}
\end{itemize}

\item Exotic Higgs boson searches
\begin{itemize}
\item Higgs boson decays to long-lived particles~\cite{ll_atlas}
\item Higgs boson decays to electron jets~\cite{ejet_atlas}
\item Higgs boson decays to displaced muon jets~\cite{mujet_atlas}
\end{itemize}

\end{itemize}
%
All searches have in common that no significant deviation from the SM expectation has been found and 
that the phase space for BSM Higgs bosons has been constrained. However, the higher kinematic 
reach at $\sqrt{s}=13-14$ TeV and the extension of the current data set by 1-2 orders of magnitude 
will allow to explore a magnitude 
scenarios over the next few years.

\section{Prospects for Higgs boson searches}
The present LHC programme projects the delivery of about 300 fb$^{-1}$ of data at $\sqrt{s}=13-14$ TeV 
per experiment by the year 2022. One of the options is that this will be followed by a high-luminosity 
LHC (HL-LHC), producing 3000 fb$^{-1}$ of data at $\sqrt{s}=14$ TeV until 2035~\cite{prosp_atlas}. In the following, projections 
of Higgs boson property measurements assuming these conditions are presented. Typically, it is assumed that the 
detector and reconstruction performance is comparable to that of LHC Run 1 due to improved reconstruction 
algorithms and detector upgrades countering the effect of an increase in concurrent proton-proton 
interactions (``pile-up''). Systematic uncertainties are assumed to be similar to the current 
status (CMS Scenario 1 and ATLAS) or to go down (CMS Scenario 2; by a factor 0.5 for theoretical uncertainties, 
and by the inverse of the square root of the integrated luminosity for other systematic uncertainties)~\cite{prosp_atlas,prosp_cms}.
\subsection{Signal strength and coupling strength projections}
\begin{figure}[htb]
\begin{center}
\begin{minipage}{0.57\textwidth}
\includegraphics[height=7.0cm]{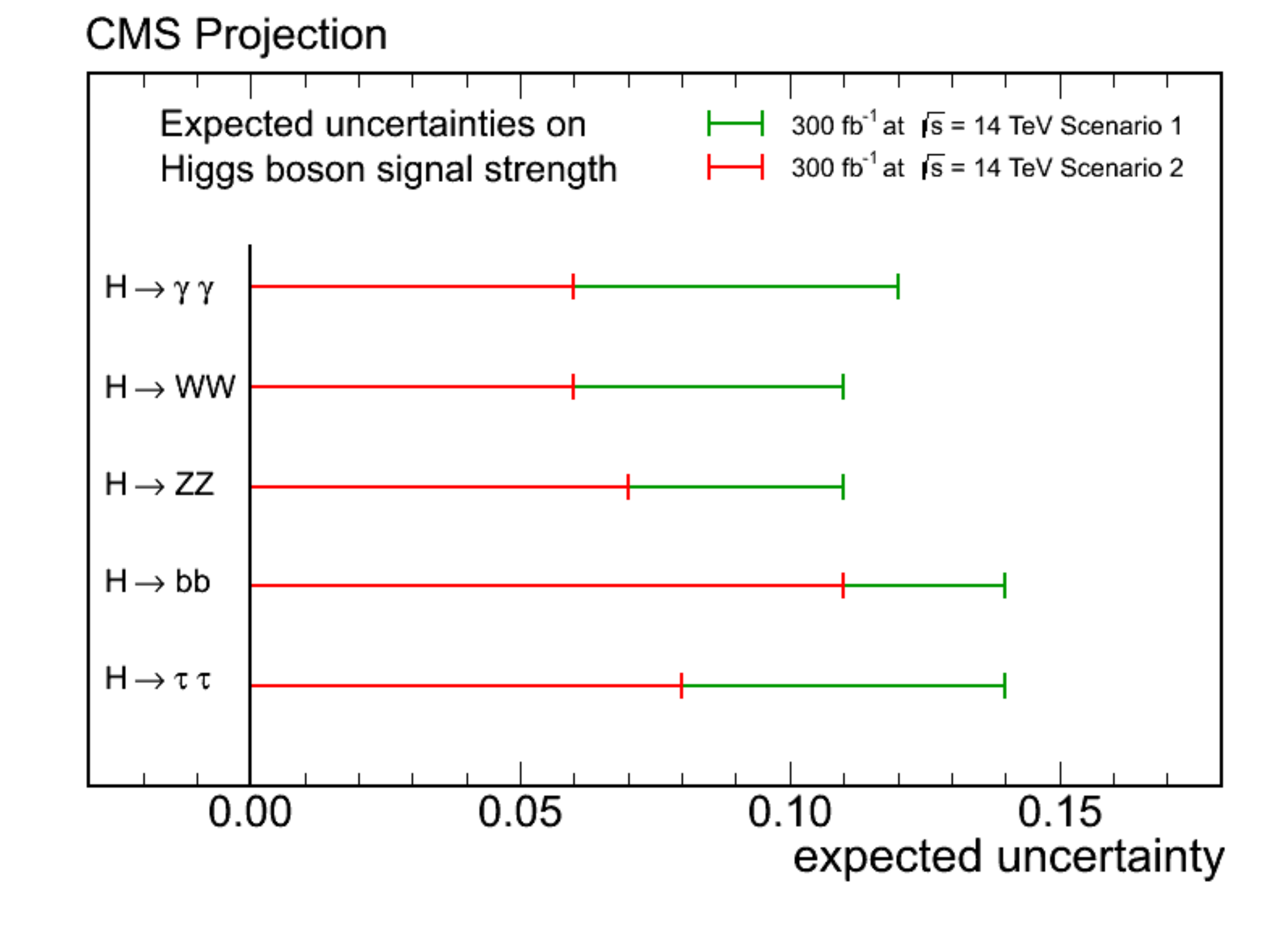}
\caption{\label{prosp_c1}CMS projection for the precision on Higgs boson signal strength measurements with 300 fb$^{-1}$~\cite{prosp_cms}. The 
two scenarios are explained in the text.}
\end{minipage}\hspace{0.04\textwidth}%
\begin{minipage}{0.35\textwidth}
\hspace{0.7cm}
\includegraphics[height=7.0cm]{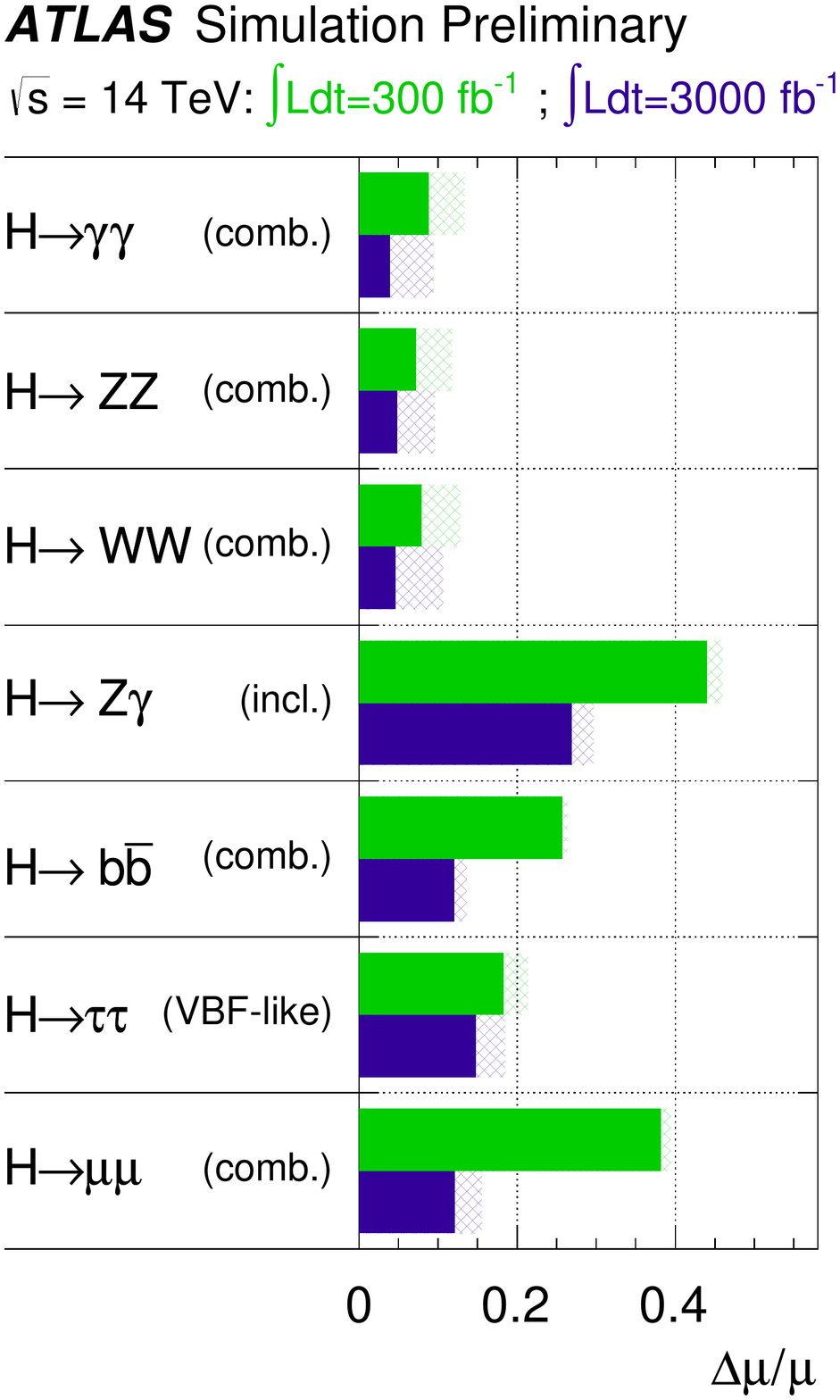}
\caption{\label{prosp_a1}ATLAS projection for the precision on Higgs boson signal strength measurements with 300 fb$^{-1}$ and 3000 fb$^{-1}$~\cite{prosp_atlas}. 
The hatched area is the effect of theoretical uncertainties.}
\end{minipage}
\end{center}
\end{figure}
The CMS estimate for the first 300 fb$^{-1}$ is a precision of roughly 10\% on the signal strength 
for the five most sensitive channels, see Fig.~\ref{prosp_c1}. The 
predicted uncertainties are between 6\%--14\%, with lower values for the bosonic decay modes 
and higher ones for the fermionic modes, and 
depending on the scenario~\cite{prosp_cms}. 
The ATLAS projection for 300 fb$^{-1}$ is slightly more conservative, expecting a precision of 
about 10\% for the bosonic modes, and about 20\% for the fermionic modes. In 
addition, a precision of about 40\% on the rare decay channels $Z\gamma$ and $\mu\mu$ 
is estimated, see Fig.~\ref{prosp_a1}. Sizable improvements with 3000 fb$^{-1}$ are in particular expected for the $bb$, $Z\gamma$ and $\mu\mu$ 
measurements but the precision for all signal strength measurements is expected to improve significantly~\cite{prosp_atlas}.

\begin{figure}[htb]
\begin{center}
\begin{minipage}{0.45\textwidth}
\includegraphics[height=5.5cm]{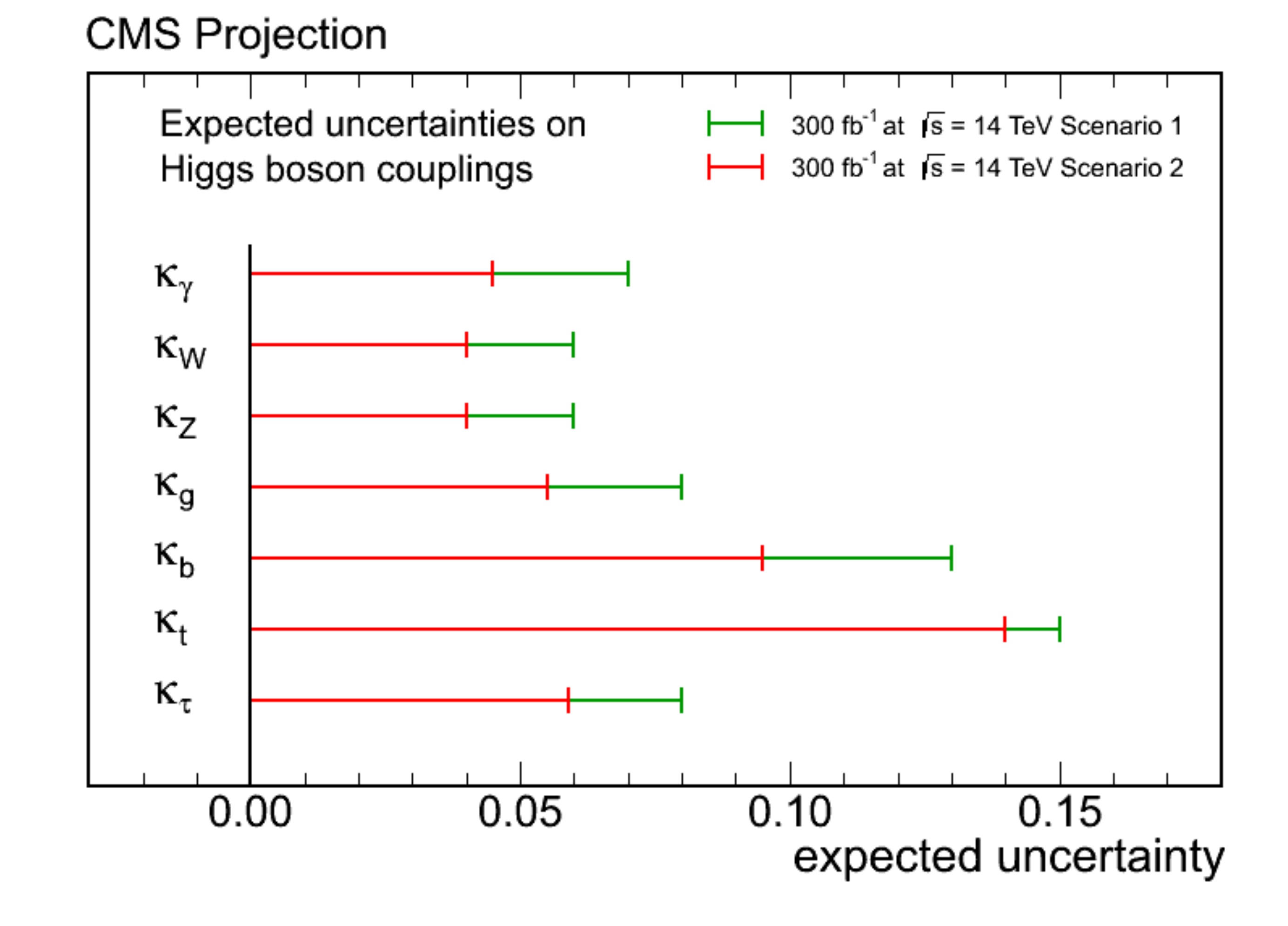}
\caption{\label{prosp_c2}CMS projection for the precision on the Higgs boson coupling strength measurements with 300 fb$^{-1}$~\cite{prosp_cms}.}
\end{minipage}\hspace{0.06\textwidth}%
\begin{minipage}{0.47\textwidth}
\includegraphics[height=4.8cm]{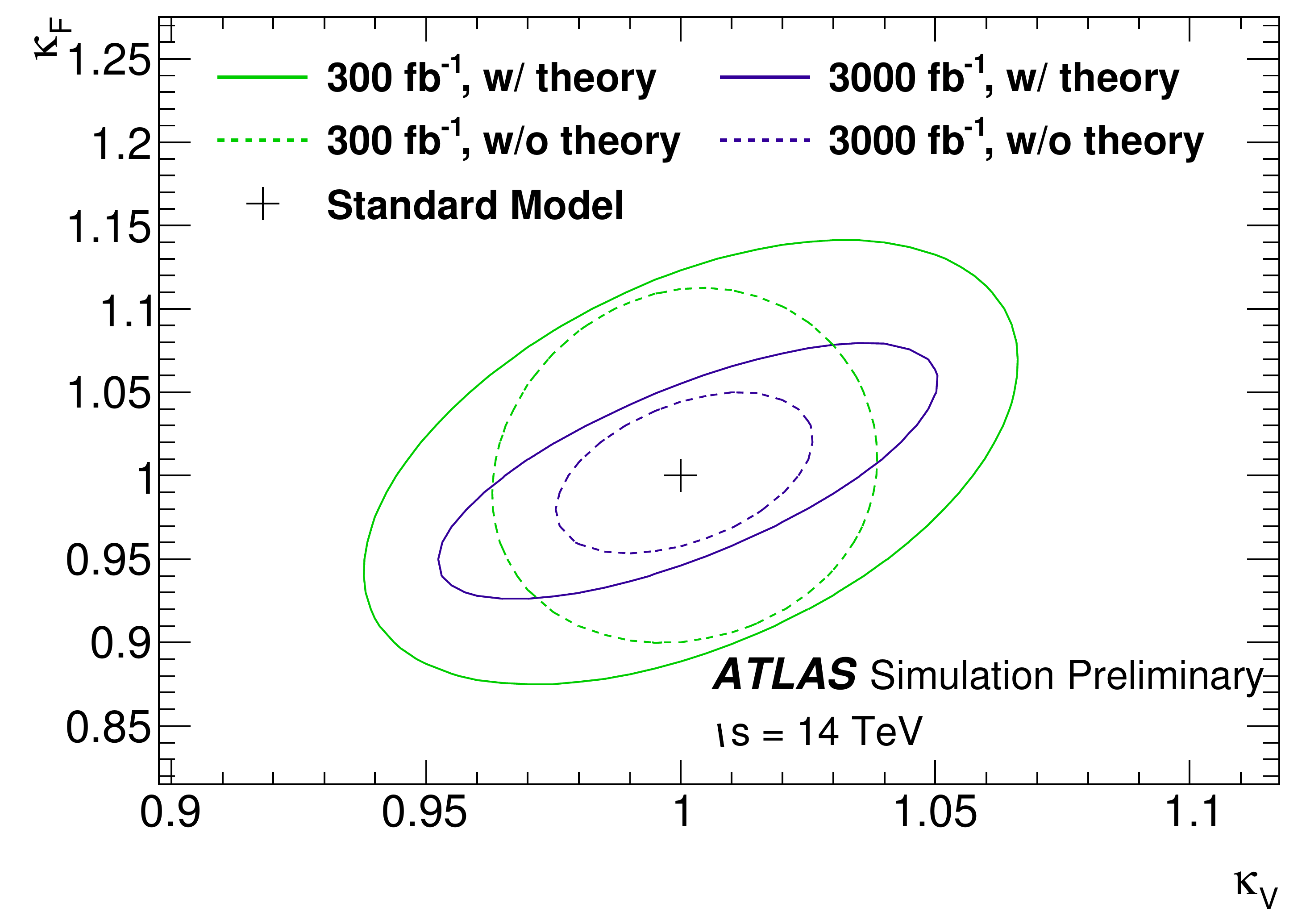}
\caption{\label{prosp_a2}ATLAS projection for the precision on the measurement of $\kappa_V$ versus $\kappa_F$ with 300~fb$^{-1}$ and 3000 fb$^{-1}$. 
The precision corresponds to a confidence level of 68\%~\cite{prosp_atlas}.}
\end{minipage}
\end{center}
\end{figure}
CMS estimates a precision of 4\%--8\% on the Higgs boson coupling strength to 
elementary bosons and the $\tau$ lepton, and of 10\%--15\% to the bottom and the top quark 
with 300 fb$^{-1}$~\cite{prosp_cms}, as illustrated in Fig.~\ref{prosp_c2}.
The ATLAS estimates are comparable but in all cases slightly more pessimistic~\cite{prosp_atlas}. 
In Fig.~\ref{prosp_a2}, the projected uncertainty for a fit with two free parameters is shown: 
$\kappa_V$ and $\kappa_F$. Assuming only these two universal scaling factors,
a precision of about 2\%--4\% on $\kappa_V$ (when $\kappa_F$ is profiled) and of 3\%--9\% on $\kappa_F$ (when $\kappa_V$ is profiled) can be achieved, 
depending on the integrated luminosity and the assumptions on theory uncertainties. The 
potential improvement on the precision of $\kappa_F$ with the HL-LHC is almost a factor of two.

\subsection{Double-Higgs production and Higgs self-coupling projections}
There are two kinds of contributions to double-Higgs production: With, and without a triple-Higgs vertex. 
A measurement of the cross section of this process thus allows to draw conclusions about the Higgs self-coupling. 
Due to the negative interference of these two contributions in the SM, only about 10 signal events are expected 
in the final state $HH \to bb\gamma\gamma$ in 3000 fb$^{-1}$ of data. 
According to ATLAS projections,
the expected significance for $HH$ production in this channel for the whole HL-LHC data set is about $1.3 \sigma$~\cite{prosp_pair_atlas}. CMS estimates
a relative uncertainty on the $HH$ cross section measurement of about 60\%~\cite{prosp_pair_cms} for the same final state.
As of today it is thus unclear if any meaningful measurement of the Higgs self-coupling at the LHC is possible.

\section{Conclusions}
Both ATLAS and CMS offer a rich Higgs physics programme. Following the Higgs boson discovery in 2012, recent years have seen a focus on 
measuring the properties of this Higgs boson. All measurements so far are consistent with the SM expectation but only the sizable improvements 
on their precision expected in the near future will allow to distinguish between the SM and many of its alternatives. In searches for BSM 
Higgs bosons, no evidence for a new state has been found, severely constraining a wide range of BSM models.

\section*{References}
\bibliography{FLECHL_martin_DISCRETE2014}

\providecommand{\newblock}{}
\begin{thebibliography}{10}
\expandafter\ifx\csname url\endcsname\relax
  \def\url#1{{\tt #1}}\fi
\expandafter\ifx\csname urlprefix\endcsname\relax\def\urlprefix{URL }\fi
\providecommand{\eprint}[2][]{\url{#2}}

\bibitem{atlas}
{ATLAS Collaboration} {\em JINST\/} {\bf 3} S08003

\bibitem{cms}
{CMS Collaboration} {\em JINST\/} {\bf 3} S08004

\bibitem{disc_atlas}
{ATLAS Collaboration} {\em Phys.Lett.\/} {\bf B716} 1--29 (\textit{Preprint}
  \eprint{1207.7214})

\bibitem{disc_cms}
{CMS Collaboration} {\em Phys.Lett.\/} {\bf B716} 30--61 (\textit{Preprint}
  \eprint{1207.7235})

\bibitem{yr1}
{LHC Higgs Cross Section Working Group}, Dittmaier S, Mariotti C, Passarino G
  and Tanaka R~E {\em CERN-2011-002\/} (\textit{Preprint} \eprint{1101.0593})

\bibitem{yr2}
{LHC Higgs Cross Section Working Group}, Dittmaier S, Mariotti C, Passarino G
  and Tanaka R~E {\em CERN-2012-002\/} (\textit{Preprint} \eprint{1201.3084})

\bibitem{yr3}
{LHC Higgs Cross Section Working Group}, Heinemeyer S, Mariotti C, Passarino G
  and Tanaka R~E {\em CERN-2013-004\/} (\textit{Preprint} \eprint{1307.1347})

\bibitem{hgg_cms}
{CMS Collaboration} {\em Eur.Phys.J.\/} {\bf C74} 3076 (\textit{Preprint}
  \eprint{1407.0558})

\bibitem{hgg_atlas}
{ATLAS Collaboration} {\em Phys.Rev.\/} {\bf D90} 112015 (\textit{Preprint}
  \eprint{1408.7084})

\bibitem{h4l_atlas}
{ATLAS Collaboration} {\em Phys.Rev.\/} {\bf D91} 012006 (\textit{Preprint}
  \eprint{1408.5191})

\bibitem{h4l_cms}
{CMS Collaboration} {\em Phys.Rev.\/} {\bf D89} 092007 (\textit{Preprint}
  \eprint{1312.5353})

\bibitem{hww_atlas}
{ATLAS Collaboration}  (\textit{Preprint} \eprint{1412.2641})

\bibitem{hww_cms}
{{{{{{{{{{{{{{CMS Collaboration}}}}}}}}}}}}}} {\em JHEP\/} {\bf 1401} 096
  (\textit{Preprint} \eprint{1312.1129})

\bibitem{htt_cms}
{{{{{{{{{{{{{{CMS Collaboration}}}}}}}}}}}}}} {\em JHEP\/} {\bf 1405} 104
  (\textit{Preprint} \eprint{1401.5041})

\bibitem{htt_atlas}
{ATLAS Collaboration} {\em ATLAS-CONF-2014-061\/}

\bibitem{tth_cms}
{CMS Collaboration} {\em JHEP\/} {\bf 1409} 087 (\textit{Preprint}
  \eprint{1408.1682})

\bibitem{hbb_atlas}
{ATLAS Collaboration}  (\textit{Preprint} \eprint{1409.6212})

\bibitem{hbb_cms}
{{{{{{{{{{{{{{CMS Collaboration}}}}}}}}}}}}}} {\em Phys.Rev.\/} {\bf D89}
  012003 (\textit{Preprint} \eprint{1310.3687})

\bibitem{tth_atlas}
{ATLAS Collaboration} {\em ATLAS-CONF-2014-043\/}

\bibitem{mass_atlas}
{ATLAS Collaboration} {\em Phys.Rev.\/} {\bf D90} 052004 (\textit{Preprint}
  \eprint{1406.3827})

\bibitem{mass_cms}
{CMS Collaboration} {\em CMS-PAS-HIG-14-009\/}

\bibitem{width_cms}
{CMS Collaboration} {\em Phys.Lett.\/} {\bf B736} 64 (\textit{Preprint}
  \eprint{1405.3455})

\bibitem{width_atlas}
{ATLAS Collaboration} {\em ATLAS-CONF-2014-042\/}

\bibitem{mu_atlas}
{ATLAS Collaboration} {Summary plots from the ATLAS Higgs physics group
  10/2014}
  \urlprefix\url{https://atlas.web.cern.ch/Atlas/GROUPS/PHYSICS/CombinedSummar%
yPlots/HIGGS/}

\bibitem{kappa_atlas}
{ATLAS Collaboration} {\em Phys.Lett.\/} {\bf B726} 88--119 (\textit{Preprint}
  \eprint{1307.1427})

\bibitem{cp_cms}
{CMS Collaboration}  (\textit{Preprint} \eprint{1411.3441})

\bibitem{cp_atlas}
{ATLAS Collaboration} {\em Phys.Lett.\/} {\bf B726} 120--144 (\textit{Preprint}
  \eprint{1307.1432})

\bibitem{diff_gg_atlas}
{ATLAS Collaboration} {\em JHEP\/} {\bf 1409} 112 (\textit{Preprint}
  \eprint{1407.4222})

\bibitem{diff_4l_atlas}
{ATLAS Collaboration} {\em Phys.Lett.\/} {\bf B738} 234--253 (\textit{Preprint}
  \eprint{1408.3226})

\bibitem{mssmn_cms}
{CMS Collaboration} {\em JHEP\/} {\bf 1410} 160 (\textit{Preprint}
  \eprint{1408.3316})

\bibitem{mssmc_atlas}
{ATLAS Collaboration} {\em ATLAS-CONF-2014-050\/}

\bibitem{mssm_benchmark}
Carena M, Heinemeyer S, St{\aa}l O, Wagner C and Weiglein G {\em Eur.Phys.J.\/}
  {\bf C73} 2552 (\textit{Preprint} \eprint{1302.7033})

\bibitem{cs_atlas}
{ATLAS Collaboration} {\em Eur.Phys.J.\/} {\bf C73} 2465 (\textit{Preprint}
  \eprint{1302.3694})

\bibitem{cs_cms}
{{{{{{{{{CMS Collaboration}}}}}}}}} {\em CMS-PAS-HIG-13-035\/}

\bibitem{tb_cms}
{{{{{{{CMS Collaboration}}}}}}} {\em CMS-PAS-HIG-13-026\/}

\bibitem{mssm_mumu_atlas}
{ATLAS Collaboration} {\em ATLAS-CONF-2012-094\/}

\bibitem{mssm_mumu_cms}
{CMS Collaboration} {\em CMS-PAS-HIG-12-011\/}

\bibitem{mssm_bb_cms}
{{{CMS Collaboration}}} {\em Phys.Lett.\/} {\bf B722} 207--232
  (\textit{Preprint} \eprint{1302.2892})

\bibitem{2hdm_ww_atlas}
{ATLAS Collaboration} {\em ATLAS-CONF-2013-027\/}

\bibitem{agg_atlas}
{ATLAS Collaboration} {\em ATLAS-CONF-2012-079\/}

\bibitem{amm_atlas}
{ATLAS Collaboration} {\em ATLAS-CONF-2011-020\/}

\bibitem{amm_cms}
{CMS Collaboration} {\em Phys.Rev.Lett.\/} {\bf 109} 121801 (\textit{Preprint}
  \eprint{1206.6326})

\bibitem{heavy_atlas}
{ATLAS Collaboration} {\em ATLAS-CONF-2013-067\/}

\bibitem{heavy_cms}
{{{{{CMS Collaboration}}}}} {\em Phys.Rev.\/} {\bf D90} 112013
  (\textit{Preprint} \eprint{1410.2751})

\bibitem{inv_atlas}
{ATLAS Collaboration} {\em Phys.Rev.Lett.\/} {\bf 112} 201802
  (\textit{Preprint} \eprint{1402.3244})

\bibitem{inv_cms}
{{{{{CMS Collaboration}}}}} {\em Eur.Phys.J.\/} {\bf C74} 2980
  (\textit{Preprint} \eprint{1404.1344})

\bibitem{hpp_atlas}
{ATLAS Collaboration}  (\textit{Preprint} \eprint{1412.0237})

\bibitem{hpp_cms}
{{{{{CMS Collaboration}}}}} {\em Eur.Phys.J.\/} {\bf C72} 2189
  (\textit{Preprint} \eprint{1207.2666})

\bibitem{lfv_cms}
{{{{{{{CMS Collaboration}}}}}}} {\em CMS-PAS-HIG-14-005\/}

\bibitem{fcnc_atlas}
{ATLAS Collaboration} {\em JHEP\/} {\bf 1406} 008 (\textit{Preprint}
  \eprint{1403.6293})

\bibitem{fcnc_cms}
{{{{{{CMS Collaboration}}}}}} {\em CMS-PAS-HIG-13-034\/}

\bibitem{fp_atlas}
{ATLAS Collaboration} {\em Eur.Phys.J.\/} {\bf C72} 2157 (\textit{Preprint}
  \eprint{1205.0701})

\bibitem{fp_cms}
{{{{{CMS Collaboration}}}}} {\em Phys.Lett.\/} {\bf B725} 36--59
  (\textit{Preprint} \eprint{1302.1764})

\bibitem{4g_atlas}
{ATLAS Collaboration} {\em ATLAS-CONF-2011-135\/}

\bibitem{bsm_coup_atlas}
{ATLAS Collaboration} {\em ATLAS-CONF-2014-010\/}

\bibitem{pair_atlas}
{ATLAS Collaboration}  (\textit{Preprint} \eprint{1406.5053})

\bibitem{pair_atlas2}
{ATLAS Collaboration} {\em ATLAS-CONF-2014-005\/}

\bibitem{pair_cms}
{{{{{{{CMS Collaboration}}}}}}} {\em CMS-PAS-HIG-14-013\/}

\bibitem{pair2_cms}
{{{{{{{CMS Collaboration}}}}}}} {\em CMS-PAS-HIG-13-032\/}

\bibitem{aa_cms}
{{{{CMS Collaboration}}}} {\em CMS-PAS-HIG-13-010\/}

\bibitem{azh_cms}
{{{{{{{CMS Collaboration}}}}}}} {\em CMS-PAS-HIG-14-011\/}

\bibitem{casc_atlas}
{ATLAS Collaboration} {\em Phys.Rev.\/} {\bf D89} 032002 (\textit{Preprint}
  \eprint{1312.1956})

\bibitem{ll_atlas}
{ATLAS Collaboration} {\em Phys.Rev.Lett.\/} {\bf 108} 251801
  (\textit{Preprint} \eprint{1203.1303})

\bibitem{ejet_atlas}
{ATLAS Collaboration} {\em New J.Phys.\/} {\bf 15} 043009 (\textit{Preprint}
  \eprint{1302.4403})

\bibitem{mujet_atlas}
{ATLAS Collaboration} {\em Phys.Lett.\/} {\bf B721} 32--50 (\textit{Preprint}
  \eprint{1210.0435})

\bibitem{prosp_atlas}
{ATLAS Collaboration} {\em ATL-PHYS-PUB-2014-016\/}

\bibitem{prosp_cms}
{CMS Collaboration} {\em CMS-NOTE-13-002\/} (\textit{Preprint}
  \eprint{1307.7135})

\bibitem{prosp_pair_atlas}
{ATLAS Collaboration} {\em ATL-PHYS-PUB-2014-019\/}

\bibitem{prosp_pair_cms}
{CMS Collaboration} {Physics Performance for 2nd ECFA workshop 10/2014}
  \urlprefix\url{https://twiki.cern.ch/twiki/pub/CMSPublic/PhysicsResultsFP/EC%
FA-CMSPublicResults.pdf}

\end{thebibliography}


\end{document}